\documentclass[ss]{imsart}

\input{macros} 

\doi{10.1214/154957804100000000}
\pubyear{0000}
\volume{0}
\firstpage{1}
\lastpage{1}

\startlocaldefs

\endlocaldefs
\usepackage{physics}

\begin{document}

\begin{frontmatter}

\title{Kronecker-structured Covariance Models for Multiway Data\support{This work was supported in part by the SOLSTICE Drive Center funded by NASA and National Science Foundation under grant 80NSSC20K0600, the Army Research Office under grants W911NF1910269 and W911NF1510479, and by the National Nuclear Security Administration under grant DE-NA0003921. }}
\runtitle{Multiway covariance models}

\begin{aug}
\author{\fnms{Yu} \snm{Wang}
\ead[label=e1]
{wayneyw@umich.edu}},
\author{\fnms{Zeyu} \snm{Sun}
\ead[label=e2]
{zeyusun@umich.edu}},
\author{\fnms{Dogyoon} \snm{Song}
\ead[label=e3]
{dogyoons@umich.edu}} 
\and
\author{\fnms{Alfred} \snm{Hero}
\ead[label=e4]
{hero@umich.edu}}

\address{University of Michigan\\ Ann Arbor, MI 48109\\
\printead{e1,e2,e3,e4}}



\runauthor{Wang et al.}

\affiliation{University of Michigan}

\end{aug}

\begin{abstract}
Many applications produce multiway data of exceedingly high dimension.
Modeling such multi-way data is important in multichannel signal and video
processing where sensors produce multi-indexed data, e.g. over spatial,
frequency, and temporal dimensions. We will address the challenges of
covariance representation of multiway data and review some of the
progress in statistical modeling of multiway covariance over the past two
decades, focusing on tensor-valued covariance models and their inference. We
will illustrate through a space weather application: predicting the evolution of solar active regions over time. 
\end{abstract}

\begin{keyword}[class=MSC]
\kwd[Primary ]{62H12}
\kwd{60H15}
\kwd{60H20}
\kwd[; secondary ]{62P12}
\end{keyword}

\begin{keyword}
\kwd{Tensor valued data} \kwd{multiway graphical lasso} \kwd{high dimensional statistics} \kwd{space weather applications}

\end{keyword}


\tableofcontents

\end{frontmatter}

\section{Introduction}\label{sec:intro}

Probabilistic modeling of data is ubiquitous in statistics and machine learning, and arguably, estimating dependencies across covariates is one of the most fundamental modeling tasks. 
While first order ``mean field" analysis can describe typical  mean behavior of covariates, higher order analysis is required to capture interactions between covariates. Among the latter, covariance estimation is a second order analysis that 
has been extensively studied 
for empirically estimating the covariance or the inverse covariance (precision) matrix from multivariate data. Besides being of interest in their own right, covariance estimates are often a prerequisite in other applications, including: dimensionality reduction; multivariate clustering; supervised classification; linear prediction; and model selection \cite{jolliffe1986principal, meinshausen2006high, yuan2007model, banerjee2008model, friedman2008sparse, tan2022rise, tan2022identifying, tan2022doubly}. 

A striking feature of many modern datasets is their high-dimensionality and multiway nature, often involving a huge number of multi-indexed variables represented as tensor-valued data \cite{kolda2009tensor, landsberg2011tensors}. 
Second order analysis of such tensor-valued data poses significant modeling and computational challenges due to the intrinsically high computational complexity of manipulating tensor representations \cite{hillar2013most}. To overcome these challenges, several methodological approaches have been developed over the past 50 years that reduce model complexity by imposing sparse or low-dimensional dependency structures in the covariance or inverse covariance.
%
These include Kronecker generalizations of PCA (Kronecker PCA) \cite{tsiligkaridis2013covariance,greenewald2014kronecker} and the matrix normal model \cite{dawid1981some}, where the covariates are multivariate normal with multiway covariance structure modeled as a Kronecker product of matrices of much lower dimension. These Kronecker  representations of covariance can have Kronecker factors that are either dense  \cite{dutilleul1999mle, werner2008estimation} or sparse \cite{allen2010transposable, tsiligkaridis2013convergence}. 

The introduction of such parsimoniously   structured tensor covariance models, and fast iterative computational algorithms for inferring these models, have made tensor covariance models more practical for many applications. 
In particular, such models have been applied to image classification \cite{fu2008image}, spatio-temporal image processing \cite{greenewald2014kronecker, greenewald2016robust,mo2022point3d,deng2021correlation}, bioinformatics \cite{teng2009statistical,greenewald2015robust,tan2022tree,du2017cbinderdb}, atmospheric science \cite{li2008three, greenewald2019tensor}, neuroscience \cite{wang2020sylvester,tan2018changepoint,bijma2005spatiotemporal},
medical imaging \cite{llosa2020reduced,liu2019regional}, and digital advertising \cite{hao2021sparse}. Several other applications of multiway covariance models are discussed in the unpublished survey \cite{suntensors}.

In this article, we survey some of the recent advances in second order analysis of multiway data. 
The focus is on Kronecker structured representations and approximations to multiway covariance. 
After introducing some notation in Section \ref{sec:prelim}, we describe how multiway covariance structure arises in different applications, including in linear prediction of dynamic processes whose evolution is governed by differential equations. 
We then overview recent approaches to Kronecker modeling and estimation of multiway covariance and inverse covariance in Section \ref{sec:models}. 
In  Section \ref{sec:performance}, we summarize and compare these Kronecker models and estimators based on their theoretical performance properties and a numerical comparison study. 
The numerical study focuses on model comparison for an important application in the domain of space weather: prediction of solar active region evolution over time.
In Section \ref{sec:conclusion}, we conclude the paper with a few open research directions in second order multiway data modeling.
A software package~\citep{WANG2022100308}
accompanies this survey with code implementing the methods described herein.


This survey is intended to complement previous surveys on aspects of high dimensional covariance modeling important to practitioners. These include surveys on robust structured covariance estimation  \cite{wiesel2015structured,ke2019user} and regularized covariance estimation \cite{pourahmadi2011covariance,ledoit2020power,chen2011robust}.  Here, we focus on the class of Kronecker structured covariance models.

\section{Background}
\label{sec:prelim}
\subsection{Notation}

Throughout the paper, scalars are denoted by lowercase letters, vectors by boldface lowercase letters, and matrices by boldface capital letters. All matrix and vector quantities are assumed to be real valued, although all the multiway covariance models discussed in this paper straightforwardly generalize to complex valued data. For a matrix $\mat{A} = (\mat{A}_{i,j}) \in \mathbb{R}^{d \times d}$, the spectral norm is denoted as $\|\mat{A}\|_2$, the Frobenius norm as $\|\mat{A}\|_F$, and the nuclear norm as $\|\mat{A}\|_*$. We define $\|\mat{A}\|_{1,\text{off}} := \sum_{i \neq j} |\mat{A}_{i,j}|$ as its off-diagonal $\ell_1$ norm. For tensor algebra, we adopt the notations used by \citet{kolda2009tensor}. $K$-th order tensors are denoted by boldface Euler script letters, e.g, $\tensor{X} \in \Reals^{d_1 \times \dots \times d_K}$, with the $(i_1,\dots, i_K)$-th element denoted as $\tensor{X}_{i_1,\dots, i_K}$. A fiber of a tensor is a higher order analogue of a row or a column of a matrix. It is obtained by fixing all but one of the indices of the tensor. The vectorization of $\tensor{X}$, denoted as $\vecto(\tensor{X})$, stacks the entries of the tensor into a $d$-dimensional column vector $(\tensor{X}_{1,1,\dots,1},\tensor{X}_{2,1,\dots,1},\dots,\tensor{X}_{d_1,d_2,\dots,d_k})^T$, where $d=\prod_{k=1}^K d_k$. Matricization, also known as unfolding, is the process of transforming a tensor into a matrix. The mode-$k$ matricization of a tensor $\tensor{X}$, denoted by $\tensor{X}_{(k)}$, arranges the mode-$k$ fibers to be the columns of the resulting matrix. The $k$-mode product of a tensor $\tensor{X} \in \Reals^{d_1 \times \dots \times d_K}$ and a matrix $\mat{A} \in \Reals^{J \times d_k}$, denoted as $\tensor{X} \times_k \mat{A}$, is a tensor of size $d_1 \times \dots \times d_{k-1} \times J \times d_{k+1} \times \dots d_K$, with entries defined as $(\tensor{X} \times_k \mat{A})_{i_1,\dots,i_{k-1},j,i_{k+1},\dots,i_K} := \sum_{i_k=1}^{d_k} \tensor{X}_{i_1,\dots,i_K} A_{j,i_k}$.
We define the $K$-way Kronecker product (KP) as $\bigotimes_{k=1}^K \mat{A}_k = \mat{A}_1 \otimes \cdots \otimes \mat{A}_K$, and the 
$K$-way Kronecker sum (KS) as $\bigoplus_{k=1}^K \mat{A}_k = \mat{A}_1 \oplus \dots \oplus \mat{A}_K = \sum_{k=1}^K \mat I_{[d_{1:k-1}]} \otimes \mat{A}_k \otimes \mat I_{[d_{k+1:K}]}$, where $\mat I_{[d_{k:\ell}]} = \mat I_{d_k} \otimes \dots \otimes \mat I_{d_\ell}$. For the case of $K=2$, $\mat{A}_1 \oplus \mat{A}_2 = \mat{A}_1 \otimes \mat{I}_{d_2} + \mat{I}_{d_1} \otimes \mat{A}_2$. 

Statistical convergence rate will be denoted by the $O_P(\cdot)$ notation, which is defined as follows. Consider a sequence of real random variables $\{X_n\}_{n \in \mathbb{N}}$ defined on a probability space $(\Omega, \mathcal{F}, P)$ and a deterministic (positive) sequence of reals $\{b_n\}_{n \in \mathbb{N}}$. By $X_n = O_P(1)$ is meant: $\sup_{n \in \mathbb{N}} P(|X_n| > K) \rightarrow 0$ as $K \rightarrow \infty$, where $X_n$ is a sequence indexed by $n$. The notation $X_n = O_P(b_n)$ is equivalent to $\frac{X_n}{b_n} = O_P(1)$. By $X_n = o_p(1)$ is meant $P(|X_n| > \epsilon) \rightarrow 0$ as $n \rightarrow \infty$ for any $\epsilon > 0$. By $\lambda_n \asymp b_n$ is meant $c_1 \leq \frac{\lambda_n}{b_n} \leq c_2$ for all $n$, where $c_1, c_2 > 0$ are absolute constants. The asymptotic notation $a_n = O(b_n)$ means $\lim\sum_{b \rightarrow \infty} |\frac{a_n}{b_n}| \leq C$ for some constant $C > 0$, while $c_n = \Omega(d_n)$ means $\lim\inf_{n \rightarrow \infty} |\frac{c_n}{d_n}| \geq C'$ for some constant $C' > 0$.

\subsection{Multiway Data}
\paragraph{Multiway data frame}
Multiway data are data that are indexed over multiple domains, called modes, and can be represented as a  tensor-valued data frame. The number of modes is called the order of the tensor. Examples of multiway data include 3D images of the brain, where the modes are the 3 spatial dimensions,  and spatio-temporal weather imaging data, a set of image sequences represented as 2 spatial modes and 1 temporal mode. One can also obtain a multiway data by stacking independent, identically distributed data samples.

\begin{figure}[!tbh]
    \centering
    \includegraphics[width=\linewidth]{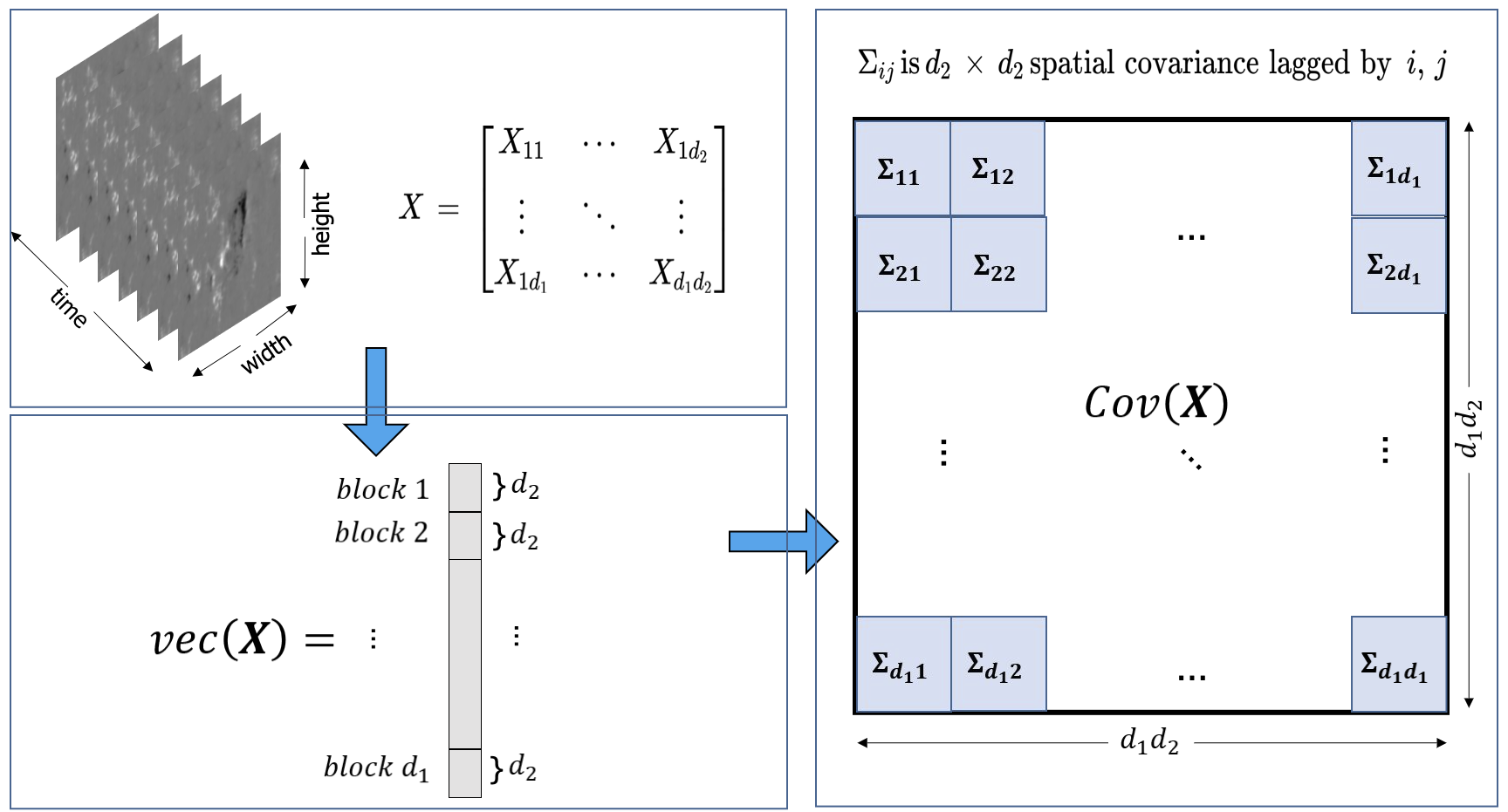}
    \caption{Multiway data and patterned covariance.}
    \label{fig:patternedcov}
\end{figure}

\paragraph{Multiway patterned covariance matrix}
Given a multiway data tensor of order $K$, the covariance matrix of the vectorized
data has a special patterned structure. For instance, suppose that $\mat{X} \in \mathbb{R}^{d_1 \times d_2}$ is a multiway data frame of order $2$, (i.e., a matrix). Then the covariance matrix of $\vecto(\mat{X})$ is a $d_1 d_2$ by $d_1 d_2$ matrix 
$\mat{\Sigma}=(\Sigma_{ij})$ that corresponds to the covariance of the $i$-th and the $j$-th rows of $\mat{X}$. See Figure \ref{fig:patternedcov} for a visual illustration.
As a special case, if the rows of $\mat{X}$ are independent and identically distributed with $d_2 \times d_2$ covariance $\mat{\Sigma}_{d_2}$, then $\cov(\mat{X}) = \mat{I}_{d_1} \otimes \mat{\Sigma}_{d_2}$.

\paragraph{Utility of multiway data for prediction}
Arguably, regression analysis is one of the most straightforward approaches for prediction as it attempts to estimate the relation between the covariates and the response variable. 
Despite its long and fruitful history, traditional regression techniques often fall short of handling multiway data encountered in modern applications, due to the ultrahigh data dimensionality as well as complex data structure. Recently, several tensor regression models were proposed to exploit the special structure of tensor covariates \cite{zhou2013tensor, hao2021sparse} or of tensor responses \cite{rabusseau2016low, li2017parsimonious, zhou2021partially}. These models try to utilize the sparse and low-rank structures in the tensors -- either in the regression coefficient tensor or the response tensor -- to boost 
performance on 
the regression task by reducing the number of free parameters. Similarly, in Section~\ref{subsec:ar_pred}, we demonstrate the utility of sparse precision/covariance matrices in linear prediction for tensor-valued data.

\subsection{Multiway covariance representations for diffusion processes}
\label{sec:diffusions}

While multiway representations have been successfully applied to many different types of data, they are particularly interpretable when modeling data generated from physical diffusion processes. In particular, consider a random process $u$ whose sample paths obey a partial differential equations of the form 
\begin{equation}
    \label{eq:pde}
    \begin{aligned}
        \mathcal{D}u &= f \quad \text{in } \Omega, \\
        u & = g \quad \text{on } \partial\Omega,
    \end{aligned}
\end{equation}
where $u$ is the unknown physical process, $f$ is the driving process (e.g., white Gaussian noise), $g$ is the function value of $u$ on the boundary, $\mathcal{D}$ is some differential operator (e.g, a Laplacian or an Euler-Lagrange operator), and $\Omega$ is the domain. After finite difference discretization over the domain $\Omega$, the model is equivalent to (ignoring discretization error) the matrix equation
\begin{equation*}
    \bD\bu = \mat{f}.
\end{equation*}
Here, $\bD$ is a sparse matrix since $\mathcal{D}$ is a differential operator. Additionally, as shown below, $\bD$ admits the Kronecker structure as a mixture of Kronecker sums and Kronecker products.

The matrix $\bD$ reduces to a Kronecker sum when $\mathcal{D}$ involves no mixed derivatives. As an example, we consider the Poisson equation, an elliptical PDE that governs many physical processes including electromagnetic induction, heat transfer, and convection. On a rectangular region $\Omega=(0,d_1)\times(0,d_2)$ in the 2D Cartesian plane, the Poisson equation with homogeneous Dirichlet boundary condition is expressed as
\begin{equation}
    \begin{aligned}
        \mathcal{D}u = (\partial^2_x + \partial^2_y)u &= f \quad \text{in } \Omega, \\
        u &= 0 \quad \text{on } \partial\Omega
        \label{eq:Poisson}
    \end{aligned}
\end{equation}
where $f: \Omega \to \bbR$ is the given source function and $u: \Omega \to \bbR$ is the unknown process of interest. Using the finite difference method with a square mesh grid with unit spacing, the unknown and the source can be expressed as $d_1$-by-$d_2$ matrices, $\bU$ and $\bF$, respectively, that are related to each other via
\begin{align}
    U_{i+1,j} + U_{i-1,j} + U_{i,j+1} + U_{i,j-1} - 4 U_{i,j} = F_{i,j}
\end{align}
for any interior grid point $(i,j)$. Defining $n$-by-$n$ square matrix
\begin{equation*}
\mat{A}_n = 
    \begin{bmatrix}
    2   &   -1  &       &   \\
    -1  &   2   & \ddots&   \\
        & \ddots& \ddots& -1\\
        &       &   -1  & 2
    \end{bmatrix},
\end{equation*}
the above relation can be expressed as the (vectorized) Sylvester equation with $K=2$:
\begin{equation}
    (\bA_{d_1} \oplus \bA_{d_2})\bu = \bff,
    \label{eq:poisson_discrete}
\end{equation}
where $\bu = \vecto(\bU)$, $\bff = \vecto(\bF)$. Note that $\mat{A}_n$ is tridiagonal. In the case where $\mat{f}$ is white noise with variance $\sigma^2$, the inverse covariance matrix of $\bu$ has the form $\cov^{-1}(\bu)=\sigma^{-2}(\bA_{d_1} \oplus \bA_{d_2})^T(\bA_{d_1} \oplus \bA_{d_2})$ and hence sparse.

More generally, any physical process generated from Equation \eqref{eq:pde} also has sparse inverse covariance matrices due to the sparsity of general discretized differential operators. Note that similar connections between continuous state physical processes and sparse ``discretized'' statistical models have been established by \citet{lindgren2011explicit}, who elucidated a link between Gaussian fields and Gauss Markov Random Fields via stochastic partial differential equations. 

In some applications, e.g. weather forecasting and multivariate time series, the data frame $\mat{X}$ can be modeled as having been generated by an underlying dynamical process obeying the aforementioned physical diffusion model. This model can be used for various tasks such as forecasting, classification, and two-sample comparison. For example, when the dynamical model is of known linear form, the Kalman filter can be used for these tasks \cite{box2015time}.
In the case of  non-linear dynamics the ensemble Kalman filter (EnKF) \cite{evensen1994sequential} is applicable. Recently, a sparsity-penalized EnKF was introduced for modeling non-linear state dynamics \cite{hou2021penalized}. This EnKF was implemented with a sparse inverse covariance estimator in order to stabilize the error covariance update of the Kalman Filter. When the state is high dimensional and tensor-valued, \cite{wang2021multiway} used Kronecker-structured models for the (inverse) covariance and demonstrated improvements in tracking certain classes of dynamical models. A similar application of Kronecker covariance modeling will be further illustrated in Section \ref{subsec:enkf}.

\section{Multiway Covariance Models}\label{sec:models}
In this section, we survey several covariance models for multiway data and review associated  estimation algorithms that have been introduced in the literature. 
The common assumption of these models is that the multimodal covariance (or  inverse covariance) of multiway data has a factorized representation involving a Kronecker product or a Kronecker sum of smaller matrices, or as a linear combination thereof. Each of the smaller matrices corresponds to a single mode of data and these are often interpreted as mode-wise (inverse) covariances. The models and the algorithms reviewed in this section are summarized in Table \ref{tab:model}.


\begin{table}
\centering
\caption{
Overview of the models and the algorithms reviewed in Section \ref{sec:models}. 
Here, $\Sigma$ and $\Omega$ denote the covariance and the inverse covariance (precision), respectively. 
For the sake of brevity, the model in the second column is described for $K=2$.}
\label{tab:model}
\resizebox{\columnwidth}{!}{
\begin{tabular}{ l c c c}
\toprule
    & Model & Algorithm & Note \\
\midrule
  \multirow{4}{*}{\makecell[l]{KP-Covariance\\(Section \ref{sec:KP_covar})}}
  & \multirow{4}{*}{ $ \bSigma=\sum_{l=1}^r \bA_l \otimes \bB_l $ }
    &   KPCA \cite{tsiligkaridis2013covariance} & Algorithm requires $K=2$\\
    &&   Robust KPCA \cite{greenewald2015robust} & Algorithm requires $K=2$\\
    &&   AdaKron \cite{leng2018covariance} & Analysis restricted to $K=2$\\
    &&   TRCM \cite{allen2010transposable} & Analysis restricted to $K=2$\\
    &&   Gemini \cite{zhou2014gemini} & Analysis restricted to $K=2$\\
\midrule
  \multirow{2}{*}{\makecell[l]{KP-Precision\\(Section \ref{sec:KP_precision})}}
  & \multirow{2}{*}{ $ \bOmega= \bA\otimes \bB $}
    &   KGlasso \cite{allen2010transposable, tsiligkaridis2013convergence} & Analysis restricted to $K=2$\\
    &&   Tlasso \cite{lyu2019tensor} &  Applicable to $K \geq 2$ \\
\midrule
  \multirow{2}{*}{\makecell[l]{KS-Precision\\(Section \ref{sec:teralasso})}}
  & \multirow{2}{*}{ $ \bOmega= \bA \oplus \bB $}
    &   Teralasso \cite{greenewald2019tensor} & Applicable to $K \geq 2$\\
    &&   BiGlasso \cite{kalaitzis2013bigraphical} &  Analysis restricted to $K=2$ \\
\midrule
  \multirow{2}{*}{\makecell[l]{Sylvester GM\\(Section \ref{sec:sylvGM})}}
  & \multirow{2}{*}{ $ \bOmega= (\bA \oplus \bB)^2 $}
    &   SyGlasso \cite{wang2020sylvester}    & Applicable to $K \geq 2$   \\
    &&   SG-PALM \cite{wang2021sg} & Applicable to $K \geq 2$ \\
\bottomrule
\end{tabular}
}
\end{table}


\subsection{The Kronecker Product Covariance Model}
\label{sec:KP}

For a $K$-mode tensor-valued data frame, the Kronecker product (KP) model of its covariance matrix $\bSigma$ is 
\begin{equation}\label{eq:KPmodel}
    \bSigma = \bSigma_1 \otimes \dots \otimes \bSigma_K
\end{equation}
where $\bSigma_k\in \Reals^{d_k \times d_k}, k=1, \ldots, K$ are symmetric positive-semidefinite matrices, which are called the Kronecker factors of $\bSigma$.

The KP model with $K=2$ imposes a simple two-way covariance representation recast as $\bSigma=\bfA\otimes \bfB$, arguably the simplest non-trivial one. Then, under this model, each $d_2 \times d_2$ block $[\bSigma]_{ij}$ of the $d_1 d_2 \times d_1 d_2$ patterned covariance matrix $\bSigma$ shown in Figure \ref{fig:patternedcov} is the same matrix, namely $[\bSigma]_{ij}=\bfB$, up to a multiplicative factor.
Despite its apparent simplicity, the KP model with $K=2$ has been used effectively in many different applications, including spatio-temporal data frames arising in MIMO wireless communications \cite{yu2001second, werner2008estimation}, geostatistics \cite{cressie2015statistics}, genomics \cite{yin2012model,hori2016multi},
face recognition \cite{zhang2010learning}, and recommender systems \cite{allen2010transposable}.

The KP model was introduced by \citet{dawid1981some} in the context of the matrix normal distribution   for data with $K=2$ tensor modes. An alternating optimization algorithm for maximum likelihood estimation (MLE) of the $K=2$ Kronecker factors was introduced by \citet{mardia1993spatial} and analyzed by \citet{dutilleul1999mle}.  A different approach based on the covariance matching principle was taken by \citet{strobach1995low}, who cast the problem of estimating the $K=2$ Kronecker factors as a minimum Frobenius norm covariance matching problem:
\be
\min_{\bfA,\bfB}\| \bS-\bA \otimes \bB\|_F^2. \label{eq:cov_matching}
\ee
Using the fact that the matrix Frobenius norm is invariant to permutation of the entries, the above optimization (\ref{eq:cov_matching}) boils down to finding the rank-one approximation to the matrix $\calR(\bS)$, where $\calR$ is the rearrangement-permutation operator that maps the square matrix  $\bS \in \Reals^{d_1 d_2 \times d_1 d_2}$ to a rectangular matrix $\calR(\bS) \in \Reals^{d_1^2 \times d_2^2}$ \cite{van1993approximation}. The minimizers of (\ref{eq:cov_matching}) can thus be expressed in closed form using the left and right principal components $\bu$ and $\bv$ of $\calR(\bS)$ as $\widehat{\bA}=\calR^{-1}(\bu)$ and $\widehat{\bB}=\calR^{-1}(\bv)$, and the resultant estimator can be represented as $\widehat{\bSigma} =\widehat{\bA}\otimes \widehat{\bB}$. Compared to the MLE, this least square estimate does not require an iterative algorithm, but is generally not asymptotically efficient. A weighted Frobenius norm formulation based on (\ref{eq:cov_matching}) was considered by \citet{werner2008estimation}, leading to an asympotically efficient estimator under the matrix normal assumption.

KP models with $K>2$ also arise in applications and are explored in the literature. 
\citet{mardia1993spatial} considered a spatio-temporal multivariate environmental monitoring data set and adopted a KP covariance with $K=3$. They were the first to propose a maximum likelihood estimation via triple iteration. As there were insufficient replicates, they simplified the model by assuming the temporal components were statistically independent. 
\citet{galecki1994general} discussed five 3-way KP models with different combinations of covariance factor structures of identity, compound symmetry, AR(1), unstructured.
\citet{akdemir2011array} provided a heuristic method for $K\geq 2$, generalizing the flip-flop method originally proposed for matrices. The convergence of the algorithm was empirically verified using simulations with $K$ up to 4, and using real world data with $K$ up to 3. 
\citet{hoff2011separable} discussed maximum likelihood estimation and Bayesian estimation for tensor normal models for general $K$, the latter approach verified on a 4-way international trade dataset. We note here that, in both \citet{hoff2011separable} and \citet{ohlson2013multilinear}, the KP model was referred to as the ``separable covariance'' model. 
An empirical study of the three-stage iterative algorithm for MLE of KP structured covariance with $K=3$ was given by \citet{manceur2013maximum}, who provided numerical and simulation results showing convergence of the algorithm.
%
\citet{pouryazdian2016candecomp}) proposed to select the order of Canonical Decomposition (CANDECOMP) in modeling 3-way data under the assumption that the noise covariance is a KP model with $K=3$ modes, justified by the structure of  electroencephalogram (EEG) data spanning the time, frequency, and space dimensions.

\subsection{Kronecker  Covariance Decomposition and Kronecker PCA}\label{sec:KP_covar}
There is a less restrictive model than \eqref{eq:KPmodel} that approximates the covariance as a sum of several Kronecker products. For example, when $K = 2$, the covariance under this model is expressed as 
\be
\bSigma=\sum_{l=1}^r \bA_l \otimes \bB_l
 \label{eq:KPCA},
\ee
where $r \geq 1$ is a positive integer, $\bA_l \in \Reals^{d_1 \times d_1}$ and $\bB_l \in \Reals^{d_2 \times d_2}$ for all $1 \leq l \leq r$. 
The sum of Kronecker product representation (\ref{eq:KPCA}) is universal in the sense that any matrix $\bSigma \in \Reals^{d_1 d_2 \times d_1 d_2 }$ can be expressed as a sum of $r$ Kronecker products with $r\leq \min\{d_1^2,d_2^2\}$ in view of SVD, cf. Van Loan and Pitzianis \cite[Section 2]{van1993approximation}. 
Indeed, even when $K \geq 3$, every tensor $\bSigma \in \Reals^{d \times d}$ can be represented by a sum of Kronecker products $\bSigma=\sum_{l=1}^r \bA^{(1)}_l \otimes \dots \otimes \bA^{(K)}_l$ where $d = \prod_{k=1}^K$ and $\bA^{(k)}_l \in \Reals^{d_k \times d_k}$ for all $k \in [K]$ and $l \in [r]$. 
The minimum positive integer $r$ such that $\bSigma$ admits a representation of the form \eqref{eq:KPCA} is called the (tensor) rank of $\bSigma$ \cite{kolda2009tensor}; 
note that the KP model from Section \ref{sec:KP} corresponds to the special case with tensor rank $1$ with additional constraints enforcing the factor matrices to be symmetric positive semidefinite.

A model of the form \eqref{eq:KPCA} (with small $r$) allows for a compact, yet quite flexible, representation of the matrix $\bSigma$. Recently, \citet{dantas2019learning} investigated the utility of this model for hyperspectral image denoising through the lens of dictionary learning. 
We remark that they used the model \eqref{eq:KPCA} to approximate the first-order ``mean-field'' behavior of data rather than its covariance (cf. Section \ref{sec:intro}), and can be considered a higher-order analogue of low-rank matrix models.

In the context of second-order (covariance) analysis, Tsiligkaridis and Hero \cite{tsiligkaridis2013covariance} proposed a penalized optimization approach for estimating a (tensor) rank $r$ Kronecker product decomposition \eqref{eq:KPCA} of covariance, proposing an estimate that solves
$$
\min_{\{\bA_l,\bB_l\}} \left\| \bS-\sum_{l=1}^r \bA_l \otimes \bB_l \right\|^2_F+\lambda\left\|\sum_{l=1}^r \bA_l \otimes \bB_l \right\|_*
$$
for  a user-supplied regularization parameter $\lambda>0$. The solution to this penalized optimization is specified by the first $r$ principal components of the  singular value decomposition (SVD) of $\calR(\bS)$ where $r$ is determined by $\lambda$ through a soft-thresholding of the SVD spectrum. In analogy to the ordinary PCA algorithm, the soft-thresholding SVD solution to this optimization problem was called Kronecker PCA (KPCA) in \cite{greenewald2014kronecker}.  In \cite{tsiligkaridis2013covariance}, it was shown that, when the sample size $n>d_1 d_2$, KPCA provides a consistent estimator of $\bSigma$.
Bounds on statistical convergence rates were also established. 

Subsequently, regularized variants of KPCA were proposed. \citet{greenewald2014kronecker} proposed to add one pair of diagonal matrices in the representation \eqref{eq:KPCA}, namely, $(\bA_0, \bB_0)$, to provide a better fit to homogeneous noise in spatiotemporal modeling. This was further extended to a robust Kronecker PCA \cite{greenewald2015robust}, which is equipped with additional capability to handle sparse unstructured ``outlier correlations'' that do not fit the vanilla KPCA model \`a la robust PCA \cite{candes2011robust, chandrasekaran2011rank}.

While the KPCA model is flexible in the sense that every covariance tensor admits a representation \eqref{eq:KPCA} for some $r$, there is no a priori guarantee that the factors $\bA_l, \bB_l$ are symmetric, positive semidefinite (PSD) matrices. Indeed, one can see from a dimension counting argument\footnote{
The set of $d_1 d_2 \times d_1 d_2$ PSD matrices has nonempty interior in the vector space of $d_1 d_2 \times d_1 d_2$ symmetric matrices that has dimension ${d_1 d_2 + 1 \choose 2}$. Note that the tensor product of the vector space of $d_1 \times d_1$ symmetric matrices and that of $d_2 \times d_2$ symmetric matrices is the span of Kronecker products and has dimension ${d_1 + 1 \choose 2} {d_2 + 1 \choose 2}$, which is strictly smaller than ${d_1 d_2 + 1 \choose 2}$ if $d_1, d_2 \geq 2$.} 
that if $d_1, d_2 \geq 2$, then there exists a PSD matrix $\bSigma \in \Reals^{d_1 d_2 \times d_1 d_2}$ that cannot be expressed as a sum of Kronecker products of symmetric matrices $\bA_l \in \Reals^{d_1 \times d_1}$ and $\bB_l \in \Reals^{d_2 \times d_2}$. As a PSD matrix is symmetric, the aforementioned matrix $\bSigma$ cannot be expressed as a sum of Kronecker products of PSD matrices. Unless the factors $\bA_l, \bB_l$ are PSD, we may not be able to interpret them as mode-wise `principal covariance components.' 

It remains an open question how closely an arbitrary covariance $\bSigma$ can be approximated by a sum of Kronecker product of PSD matrices, or whether a typical covariance $\bSigma$ can be represented as a sum of Kronecker product of PSD matrices under reasonable data-generative models.


\subsection{Kronecker Graphical Lasso and Tensor Lasso}\label{sec:KP_precision}
As discussed in Section \ref{sec:diffusions}, physical processes often give rise to sparse covariance or sparse inverse covariance. Furthermore, when the number of independent samples, $n$, is smaller than the data dimension, $d=\prod_{i=1}^K d_i$, imposition of a sparse structure can stabilize the multi-way covariance estimates.  Leng and Pan \cite{leng2018covariance} considered the problem of estimating a sparse covariance in the KP model (\ref{eq:KPmodel}) using a sparsity-penalized moment matching approach. Under the matrix normal model, this corresponds to a sparse model on the pairwise marginal dependencies in the covariates. It is sometimes more natural to impose sparsity on the conditional dependencies, equivalent to imposing sparsity on the inverse covariance instead of on the covariance itself, as demonstrated by Yuan and Lin \cite{yuan2007model} and Banerjee \etal \cite{banerjee2008model}. A fast algorithm for sparse estimation of the inverse covariance in the matrix normal framework was proposed by  Friedman \etal \cite{friedman2008sparse}, who called the algorithm the graphical lasso (Glasso). 

Allen and Tibshirani 
\cite{allen2012inference} extended the Glasso to the  matrix normal model ($K=2$ modes) and this algorithm, which we call the Kronecker graphical lasso (KGlasso), was shown to be consistent with statistical convergence bounds in \cite{tsiligkaridis2013convergence}. The KGlasso uses an alternating flip-flop optimization on the sparsity penalized KP log-likelihood  function, similarly to the algorithm proposed by Mardia and Goodall \cite{mardia1993spatial} for the original unpenalized KP maximum likelihood problem. Besides the diffusion processes mentioned in Section~\ref{sec:diffusions}, the sparse KP inverse covariance structures also arise in classification problems where the predictors are matrix-valued~\citep{molstad2019penalized}.

The KGlasso solves the following non-convex optimization, equivalent to maximizing the penalized log likelihood, 
\be 
\min_{\bA, \bB} \trace\{\bS(\bA \otimes \bB)\} - \log\det (\bA \otimes \bB)+ \lambda_1 \| \bA\|_1+\lambda_2\|\bB\|_1
\label{eq:KGlasso}
\ee
where $\lambda_1$,$\lambda_2$ are sparsity regularization parameters applied to the $\ell_1$ norms of the Kronecker product factors $\bA$ and $\bB$. Here, the $\ell_1$ norms are often restricted to range only over the off-diagonals of $\bA$ and $\bB$. Like the Glasso, iterative optimization methods are required to solve the optimization problem \eqref{eq:KGlasso}, however, algorithmic implementation of such methods can be significantly more difficult because it is a non-convex problem.

The tensor lasso (Tlasso) introduced by Lyu \etal \cite{lyu2019tensor} is a generalization of the KGlasso to $K\geq 2$ modes.   As in the KGlasso, an alternating flip-flop maximization method is used to estimate the precision matrix factors. Consistency and a minimax property are established for the Tlasso.  As pointed out in \cite{lyu2019tensor}, normalization of the $K$ inverse covariance factors is necessary to ensure model identifiability as the tensor normal  model is invariant to reciprocal scaling and other transformations on the Kronecker factors.  

Following KGlasso and Tlasso, recent works have focused on more efficient estimation algorithms for solving the non-convex problem posed by~\eqref{eq:KGlasso} with provable guarantees. For example, instead of alternatingly applying Glasso, \citet{xu2017efficient} proposed an alternating block gradient descent algorithm that has been shown to converge linearly to the true unknown precision matrices with optimal statistical error rates. More recently, \citet{min2022fast} proposed a parallel estimation scheme as an alternative to previous alternating/cyclic approaches for estimating the $K$ inverse covariance factors. They demonstrated that the parallelized version achieves similar convergence rates as alternating block descent.
 
\subsection{Tensor Graphical Lasso}\label{sec:teralasso}
The tensor graphical lasso (TeraLasso) of Greenewald \etal \cite{greenewald2019tensor} is a $K\geq 2$ generalization of the bigraphical lasso (BiGlasso) of Kalaitzis \etal  \cite{kalaitzis2013bigraphical}. For simplicity, we confine the discussion to the simpler $K=2$ mode case. In contrast to the KGlasso/Tlasso, the TeraLasso replaces the Kronecker product representation of the inverse covariance, $\bOmega= \bA\otimes \bB$,  with a Kronecker sum representation: $\bOmega=\bA \oplus \bB=\bA \otimes \bI + \bI \otimes \bB$. The rationale of \cite{kalaitzis2013bigraphical,greenewald2019tensor} that motivates the Kronecker sum model is that, for equal number of non-zero entries in factors $\bA$ and $\bB$, it gives a sparser inverse covariance than the Kronecker product. This can be appreciated from the fact that the Kronecker sum representation of the inverse covariance corresponds to a Cartesian product graphical model instead of a direct product graphical model given by the Kronecker product representation. This KS structure has recently been studied in the context of matrix-valued time-series analysis -- \citet{jiang2021online} introduced a novel multivariate autoregressive model to infer the graph topology encoded in the coefficient matrix, which captures the sparse Granger causality dependency structures describing both the sparse spatial relationship between sensors and the multiple measurement relationship. They decompose the graph by imposing a KS structure on the coefficient matrix.

Furthermore, the Kronecker sum representation converts non-convex optimization problem in the KGlasso (\ref{eq:KGlasso}) into the convex optimization problem:
\be 
\min_{\bA, \bB} \trace\{\bS(\bA \oplus \bB)\} - \log\det (\bA \oplus \bB)+ \lambda_1 \| \bA\|_1+\lambda_2\|\bB\|_1
\label{eq:BiGlasso}
\ee
where $\lambda_1$,$\lambda_2$ are sparsity regularization parameters applied to the $\ell_1$ norms of the Kronecker sum factors $\bA$ and $\bB$. As in the KGlasso, these $\ell_1$ norms range over the off-diagonals of $\bA$ and $\bB$. This objective can be solved via either an iterative application of Glasso~\citep{kalaitzis2013bigraphical} or by a projected proximal gradient descent algorithm~\citep{greenewald2019tensor}. More recent work has also extended these estimation schemes to be more scalable and to handle non-identifiability~\citep{yoon2020eiglasso} and count data~\citep{li2022scalable}.

\subsection{Sylvester Graphical Lasso and SG-PALM}\label{sec:sylvGM}
The Sylvester graphical lasso (SyGlasso) \cite{wang2020sylvester} models the precision matrix as the square of a sparse matrix that has Kronecker sum structure. Different from KGlasso, Tlasso and TeraLasso, the Sylvester graphical model uses the Sylvester tensor equation to define a generative process for the underlying multivariate tensor data. The Sylvester tensor equation has been studied in the context of finite-difference discretization of high-dimensional elliptical partial differential equations~\citep{grasedyck2004existence, kressner2010krylov}. Consider the simplest case of $K = 2$, any solution $\bX$ to such a PDE must have the (discretized) form:
\begin{equation}\label{eqn:sylvester}
    \begin{aligned}
        \bX \bA + \bB \bX = \bZ &\Longleftrightarrow (\bA \oplus \bB) \vecto(\bX) = \vecto(\bZ),
    \end{aligned}
\end{equation} 
where $\bZ$ is the driving source on the domain, and $\bA \oplus \bB$ represents the discretized differential operators for the PDE, e.g., Laplacian, Euler-Lagrange operators, and associated coefficients. These operators are often sparse and structured. 

The Sylvester generative (SG) model~\eqref{eqn:sylvester} leads to a matrix-valued random variable $\bX$ with a precision matrix $\bOmega=(\bA \oplus \bB)^2$, given that $\bZ$ is isotropic (white) Gaussian distributed. The Sylvester generating factors $\bA$ and $\bB$ can be obtained via minimization of the penalized negative log-pseudolikelihood
\be\label{eqn:objective}
    \min_{\bA,\bB} \tr\{\bS(\bA \oplus \bB)^2\} - \log\det \Big(\diag(\bA) \oplus \diag(\bB)\Big) + \lambda_1 \| \bA\|_1 + \lambda_2\|\bB\|_1.
\ee
This differs from the true penalized Gaussian negative log-likelihood in the exclusion of off-diagonals of $\bA$ and $\bB$ in the log-determinant term. \eqref{eqn:objective} is motivated and derived directly using the Sylvester equation defined in~\eqref{eqn:sylvester}, from the perspective of solving a sparse linear system. This is a generalization and extension of the convex pseudolikelihood framework for high dimensional partial correlation estimation (CONCORD) \cite{khare2015convex} to tensor-variate variables. The pseudolikelihood function is convex and is maximized using a entrywise regression approach similar to CONCORD. It is known that inference using pseudo-likelihood is consistent and enjoys the same $\sqrt{N}$ convergence rate as the MLE in general \citep{besag1977efficiency, varin2011overview}. 


Estimation of the generating parameters of the SG model is challenging since the sparsity penalty applies to the square root factors of the precision matrix, which leads to a highly coupled likelihood function. The nodewise regression approach in SyGlasso recovers only the off-diagonal elements of each Sylvester factor. This is a deficiency in many applications where the factor-wise variances are desired. Moreover, the convergence rate of the cyclic coordinate-wise algorithm used in SyGlasso is unknown and the computational complexity of the algorithm is higher than other sparse Glasso-type procedures. To overcome these deficiencies, Wang \etal~\citep{wang2021sg} proposed a proximal alternating linearized minimization method that is more flexible and versatile, called SG-PALM, for finding the minimizer of \eqref{eqn:objective}. SG-PALM is designed to exploit structures of the coupled objective function and yields simultaneous estimates for both off-diagonal and diagonal entries. It also achieves state-of-the-art iterative convergence result (i.e., linear convergence of the optimization error) and an order of magnitude improvement in practical runtime per iteration. Additionally, SG-PALM easily adapts to non-convex regularization functions (e.g., SCAD, MCP) and enjoys similar rate of iterative convergence.


\section{Illustrations and Performance Comparisons}\label{sec:performance}
In this section, we discuss theoretical guarantees on the multiway covariance estimation procedures described in Section \ref{sec:models} and provide numerical illustrations of them. In Section \ref{sec:guarantees}, we collect and compare theoretical guarantees on the statistical consistency and computational complexity for these covariance estimation algorithms. 
Then, in Section \ref{subsec:enkf}, we provide empirical performance comparisons when the covariance / precision matrix estimation methods are used for characterizing multiway physical processes.
Finally, in Section \ref{subsec:ar_pred}, we demonstrate the use of tensor graphical models on the real-world application of solar active region evolution prediction.
The numerical studies in Section \ref{subsec:enkf} and \ref{subsec:ar_pred} were performed using the publicly available software package~\citep{WANG2022100308} developed by the authors, and the readers can explore other properties beyond those presented here. 

\subsection{Theoretical Guarantees}\label{sec:guarantees}
All of the methods described in Section \ref{sec:models} have non-asymptotic theoretical guarantees on the statistical consistency of the estimators and the rate of computational convergence, which can be converted to upper bounds on the sample- and computational complexity of the methods, respectively. 
The best known upper bounds on the estimation error (in Frobenius norm) and those on computational complexity (in number of flops per iteration) are summarized in Table \ref{tab:guarantees}. 

\paragraph{Statistical Error}
For each row of Table \ref{tab:guarantees}, the expression for statistical error applies only when the associated model is correct, i.e., there is no model mismatch (bias) between the model generating the observations and the model assumed to estimate covariance/precision. For structured covariance estimation, the more recent AdaKron algorithm achieves better statistical error (given that the true covariance matrix is sparse) than KPCA/Robust KPCA as the dependence on $d_i$'s is linear. For structured precision matrix estimation, all the structured methods enjoy significantly reduced statistical error compared to the traditional Glasso. TeraLasso outperforms Tlasso and SyGlasso/SG-PALM approximately by a factor of $d_k$ and $s_k$, respectively. Moreover, it is noteworthy that Tlasso and TeraLasso achieve single-sample convergence of the $\ell_2$ norm (operator norm) errors that are not included in the table, meaning that the error vanishes with a single sample ($N = 1$) as the dimension grows to infinity~\citep{greenewald2019tensor,lyu2019tensor}. 

\paragraph{Computational Complexity}
KPCA-based methods require expensive full SVD/eigenvalue factorization in general, which dominates their runtime complexity. Although speedup can be achieved via randomized algorithms or iterative algorithms for sparse matrices along with small $r$, the computational complexities of these improvements remain unknown for general problem instances. 

All the structured precision estimation algorithms are variants of Glasso, implemented with techniques tailored to the model assumptions for speedup. Generally speaking, the resulting complexity consists of the mode-wise complexity ($d_k^3$) and the cost of updating the objective: $dK$ for TeraLasso, $N\sum_k d_k m_k^2$ for Tlasso, and $N \sum_k \sum_{j \neq k} d_j m_j^2$ for SG-PALM. Note that AdaKron for structured covariance models uses a pre-processing step that avoids this operation involving $d$ and/or $N$ flops per-iteration, but this works only for $K=2$. The mode-wise complexity of TeraLasso is dominated by matrix inversion, which is hard to scale for general problem instances. For Tlasso/KGlasso, the mode-wise complexity is the same as that of running a Glasso-type algorithm for each mode, which could be improved by applying state-of-the-art optimization techniques developed for vector-variate Gaussian graphical models. For SG-PALM, the mode-wise operations involve only sparse-dense matrix multiplications, which could be improved to $O(d_k \cdot \textsf{nnz})$, where $\textsf{nnz}$ counts the number of non-zero elements of the sparse matrix (i.e., the estimated $\bOmega_k$ at each iteration). This could greatly reduce the computational cost for extremely sparse $\bOmega_k$, e.g., with only $O(d_k)$ non-zero elements. Further, Tlasso and SG-PALM both incur a cost of $O(N d_k m_k^2)$ for each mode-wise update. This can also be reduced to be $\approx d$ for sparse estimated $\bOmega_k$'s at each iteration. Overall, for sample-starved setting where we only have access to a handful of data samples, structured KP and KS models run similarly fast, while the Sylvester GM runs slower theoretically due to the extra and richer structures that it takes into account.




\begin{table}
\centering
\caption{Comparison of known theoretical guarantees on sample complexity (statistical error) and computational complexity. Here, $M = \max\{d_1, d_2, N\}$, $m_k = \prod_{i \neq k} d_i$ is the co-dimension of the $k$-th mode, $d = \prod_{k=1}^K d_i$, and $s_k$ characterizes the sparsity of each of the inverse covariance Kronecker factors $s_k = |\{(i,j): i \neq j, [\bm\Omega_k]_{i,j} \neq 0\}|$, $s$ is the sparsity of the full inverse covariance $s = |\{(i,j): i \neq j, \bm\Omega_{i,j} \neq 0\}|$ and $s = \sum_{k=1}^K m_k s_k$ if $\bOmega$ satisfies the Kronecker sum model.
}
\label{tab:guarantees}
\resizebox{\columnwidth}{!}{
\begin{tabular}{ l c c c}
\toprule
  Model & Algorithm & Statistical Error & Computational Complexity \\
\midrule
  \multirow{3}{*}{\makecell[l]{KP-Covariance\\(Section \ref{sec:KP_covar})}}
    &   KPCA \cite{tsiligkaridis2013covariance} & $O_P\Big(\sqrt{\frac{r(d_1^2+d_2^2+\log M)}{N}}\Big)$ & $O(d_1^3d_2^3)$\\
    &   Robust KPCA \cite{greenewald2015robust} & $O_P\Big( \sqrt{\frac{r(d_1^2+d_2^2+\log M)}{N}} \Big)$ & $O(d_1^3 d_2^3)$\\
    &   AdaKron \cite{leng2018covariance} & $O_P\Big( \sum_{i=1}^2\sqrt{\frac{(s_i + d_i) \log d_i}{N}} \Big)$ & $O(d_1^3 + d_2^3)$\\
\midrule
  \multirow{2}{*}{\makecell[l]{KP-Precision\\(Section \ref{sec:KP_precision})}}
    &   KGlasso \cite{allen2010transposable, tsiligkaridis2013convergence} & $O_P\Big(\sqrt{\frac{(d_1+d_2)\log M}{N}}\Big)$ & $O(d_1^3 + d_2^3 +Nd_1d_2)$\\
    &   Tlasso \cite{lyu2019tensor} & $O_P\Big(\sqrt{\frac{d_k(d_k + s_k)\log d_k}{N d}}\Big)$ mode-$k$
    &  $O\Big(\sum_{k=1}^K (d_k^3 + N d_k m_k^2)\Big)$ \\
\midrule
  \makecell[l]{KS-Precision\\(Section \ref{sec:teralasso})}
    &   TeraLasso \cite{greenewald2019tensor} & $O_P\Big(\sqrt{K+1}\cdot \sqrt{\frac{(d+s) \log d}{N \min_k m_k}}\Big)$ & $O(dK+\sum_{k=1}^K d_k^3)$ \\
\midrule
  \multirow{2}{*}{\makecell[l]{Sylvester GM\\(Section \ref{sec:sylvGM})}}
    &   SyGlasso \cite{wang2020sylvester}    & $O_P\Big(\sqrt{K}\cdot \max_k \sqrt{\frac{s_k d_k \log d}{N}}\Big)$ & $O\Big(d\sum_{k=1}^K (Nd_k + \sum_{j \neq k} (d_k + d_j)) \Big)$   \\
    &   SG-PALM \cite{wang2021sg} & $O_P\Big(\sqrt{K}\cdot \max_k \sqrt{\frac{s_k d_k \log d}{N}}\Big)$ & $O\Big(\sum_{k=1}^K (d_k^3 + N \sum_{j \neq k} d_jm_j^2) \Big)$ \\
\bottomrule
\end{tabular}
}
\end{table}


\subsection{Numerical Experiments with Synthetic Data}\label{subsec:enkf}
Here we compare the performance of the Kronecker covariance / precision estimation methods when used to estimate multiway physical processes, based on numerical experiments using synthetic data. To this end, we first describe two generative models that extend the spatial Poisson equation (cf. Section~\ref{sec:diffusions}) to incorporate temporal dynamics, and the resulting multiway (inverse) covariance structure in Section \ref{sec:generative}. Thereafter, in Section \ref{sec:synthetic_exp}, we illustrate the performance of the various (inverse) covariance estimation methods using synthetic data generated from the two models of spatio-temporal dynamics.

\subsubsection{Two Generative Models for Spatio-temporal Dynamics}\label{sec:generative}
We describe two dynamic models that extend the spatial Poisson equation (cf. Section~\ref{sec:diffusions}) to incorporate temporal dynamics, and the resulting multiway (inverse) covariance structure. These models will be used in our numerical experiments to generate data to compare the performance of estimation algorithms in Section \ref{sec:synthetic_exp}.

\paragraph{Poisson-AR(1) Process.}
The first extension, which we call the Poisson-AR(1) process, imposes an autoregressive temporal model of order 1 on the source function $f$ in the Poisson equation (\ref{eq:Poisson}). Specifically, we say a sequence of discretized spatial observations $\{\bU^k \in \bbR^{d_1\times d_2}\}_k$ indexed by time step $k=1,\cdots,T$ is from a Poisson-AR(1) process if
\begin{align}
    &(\mat{A}_{d_1} \oplus \mat{A}_{d_2}) \vecto(\mat{U}^k) = \vecto(\mat{Z}^k), \\
    &\vecto(\mat{Z}^k) = a \vecto(\mat{Z}^{k-1}) + \vecto(\mat{W}^k),\quad |a|<1, \label{eq:ar1}
\end{align}
where $\{\mat{W}^k \in \bbR^{d_1\times d_2}\}_k$ is spatiotemporal white noise, i.e., $W_{i,j}^k \sim \mathcal{N}(0, \sigma^2_w)$, i.i.d.
Assuming $\mat{Z}^0 = \mathbf{0}$ and defining the $T$-by-$T$ matrix
\begin{equation*}
    \mat{B} = \begin{bmatrix}
        1 & -a &  &  \\
         & 1 & \ddots  &  \\
        & & \ddots & -a \\
        & & & 1
    \end{bmatrix},
\end{equation*}
the above linear system of equations can be written as $(\mat{A}_{d_1} \oplus \mat{A}_{d_2}) \mat{U} \mat{B} = \mat{W}$, or equivalently,
\begin{align}
    \left( \mat{B}^T \otimes (\mat{A}_{d_1} \oplus \mat{A}_{d_2}) \right) \vecto(\mat{U}) = \vecto(\mat{W}),
    \label{eq:poisson_ar_discrete}
\end{align}
where $\mat{U} = [\vecto(\mat{U}^1) \vecto(\mat{U}^2) \dots \vecto(\mat{U}^T)] \in \Reals^{d_1d_2 \times T}$ and $\mat{W}$ is defined likewise. The inverse covariance of $\mat{U}$, despite having a large size of $d_1d_2T \times d_1d_2T$, is sparse and has a mixed Kronecker sum and product structure.

\paragraph{Convection-diffusion Process.} 
The second time-varying extension of the Poisson PDE model (\ref{eq:Poisson}) is based on the convection-diffusion process~\cite{chandrasekhar1943stochastic}
\begin{equation}\label{eqn:convec-diff}
    \pdv{u}{t} = \theta \sum_{i=1}^2 \pdv[2]{u}{x_i} - \epsilon \sum_{i=1}^2 \pdv{u}{x_i}.
\end{equation}
Here, $\theta > 0$ is the diffusivity; and $\epsilon \in \Reals$ is the convection velocity of the quantity along each coordinate. Note that for simplicity of discussion here, we assume these coefficients do not change with space and time (see, \citet{stocker2011introduction}, for example, for a detailed discussion). These equations are closely related to the Navier-Stokes equation commonly used in stochastic modeling for weather and climate prediction~\citep{chandrasekhar1943stochastic,stocker2011introduction}. Coupled with Maxwell's equations, these equations can be used to model magneto-hydrodynamics~\citep{roberts2006slow}, which characterize solar activities including flares.


A solution of Equation~\eqref{eqn:convec-diff} can be approximated similarly as in the Poisson equation case, through a finite difference approach. Denote the discrete spatial samples of $u(\mat{x},t)$ at time $t_k$ as a matrix $\mat{U}^k\in\bbR^{d_1 \times d_2}$. We obtain a discretized update propagating $u(\mat{x},t)$ in space and time, which locally satisfies
\begin{equation}\label{eqn:convec-diff-discrete}
\begin{aligned}
    \frac{U_{i,j}^k - U_{i,j}^{k-1}}{\Delta t} = &\ \theta \left(\frac{U_{i+1,j}^k + U_{i-1,j}^k + U_{i,j+1}^k + U_{i,j-1}^k - 4U_{i,j}^k}{h^2}\right) \\
    &- \epsilon \left(\frac{U_{i+1,j}^k - U_{i-1,j}^k + U_{i,j+1}^k - U_{i,j-1}^k}{2h}\right),
\end{aligned}
\end{equation}
where $\Delta t = t_{k+1} - t_{k}$ is the time step and $h$ is the mesh step (spatial grid spacing). 
Similarly to the Poisson-AR(1) process, in the following, we consider a ``blocked'' version of the convection-diffusion process. 

We define the first-order and second-order discretized differential operators, denote by $\mat{D}$ and $\mat{A}$, respectively: 
\begin{equation*}
\mat{D} = 
    \begin{bmatrix}
    1   &     &       &   \\
    -1  &   1   & &   \\
        & \ddots& \ddots& \\
        &       &   -1  & 1
    \end{bmatrix}, \quad
\mat{A} = 
    \begin{bmatrix}
    2   &   -1  &       &   \\
    -1  &   2   & \ddots&   \\
        & \ddots& \ddots& -1\\
        &       &   -1  & 2
    \end{bmatrix}.
\end{equation*}
Then, Equation~\eqref{eqn:convec-diff-discrete} can be written as
\begin{equation}
\begin{aligned}
    \frac{1}{\Delta t}(\mat{D} \otimes \mat{I} \otimes \mat{I}) \vecto{\mat{U}} =& \frac{\theta}{h^2}(\mat{I} \otimes \mat{A} \otimes \mat{I} + \mat{I} \otimes \mat{I} \otimes \mat{A}) \vecto{\mat{U}} \\
    &- \frac{\epsilon}{2h} (\mat{I} \otimes \mat{D} \otimes \mat{I} + \mat{I} \otimes \mat{I} \otimes \mat{D}) \vecto{\mat{U}},
\end{aligned}
\end{equation}
where $\mat{U} = [\vecto(\mat{U}^1) \vecto(\mat{U}^2) \dots \vecto(\mat{U}^T)] \in \Reals^{d_1d_2 \times T}$. Assuming the process is driven by some white noise $\mat{W}$, similarly defined as in the Poisson-AR equation, the inverse covariance of $\mat{U}$ is again sparse and has a mixed Kronecker sum and product structure.

\subsubsection{Experiments with Synthetic Data: Performance Comparisons}\label{sec:synthetic_exp}
\begin{figure}[!tbh]
    \centering
    \includegraphics[width=0.9\linewidth]{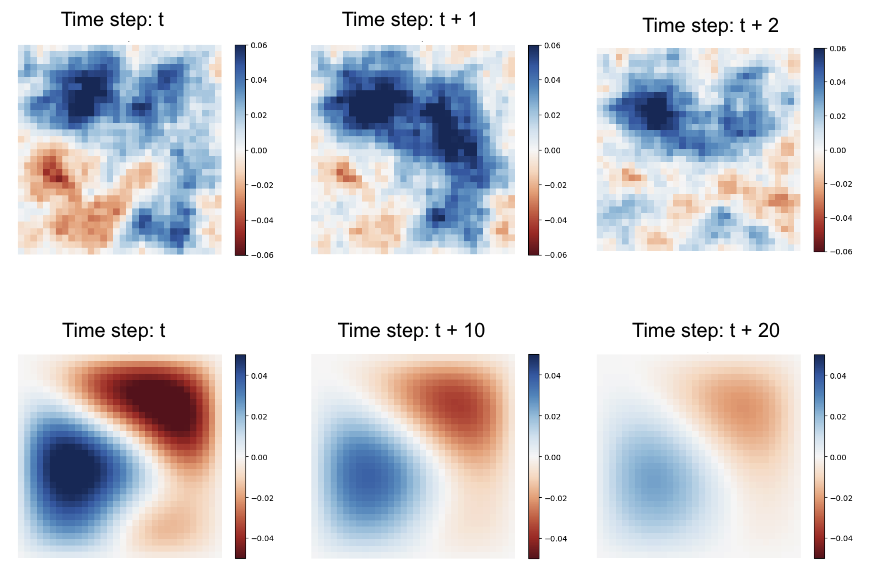}
    \caption{Poisson-AR(1) (top) and 2D Convection-diffusion (bottom) state variables of size $64 \times 64$ at three different time steps, which are chosen such that $t$ is sufficiently large for the processes to be in steady state and that the sampling times are adjusted to the temporal dynamics of each process. Here, models parameters used for simulations are: $a = -0.5$ for Poisson-AR, $\theta=0.05$ and $\epsilon=0.01$ for convection-diffusion process; the driving noise is generated from white Gaussian with $\sigma_w=0.1$ in both cases.}
    \label{fig:enkf_states}
\end{figure}

We consider spatio-temporal processes (2D space + time) that evolve $T = 50$ time steps on a $8 \times 8$ spatial grid according to the Poisson-AR(1) and the Convection-Diffusion dynamics (cf. Section \ref{sec:generative}). We generated $N = 50$ independent realizations of each type of process. Figure~\ref{fig:enkf_states} shows one realization for each type of process, with spatial resolution increased for better illustration. Assuming $K = 2$ with a spatial dimension 64 and a temporal dimension 50, we estimated the covariance / inverse covariance using several sparse multiway inverse covariance estimation methods including Glasso~\citep{friedman2008sparse}, KPCA~(Section \ref{sec:KP_covar}), 
Tlasso~(Section \ref{sec:KP_precision}), 
TeraLasso~(Section \ref{sec:teralasso}), 
SG-PALM~(Section \ref{sec:sylvGM}). Note that none of the above-mentioned models operate under the true generative processes (i.e., there is model mismatch with the data).
Here, the sparsity-regularized methods are all implemented with an $\ell_1$ penalty function, and the penalty parameters were selected similarly and guided by the theoretical results in Table~\ref{tab:guarantees}. For example, for SG-PALM, we use a penalty parameter of $\lambda_k = C \sqrt{\frac{d_k \log d}{N}}$ where $C$ is chosen by optimizing a normalized Frobenius norm error between the estimate and the truth, over a range of $\lambda$ values parameterized by $C$. For the KPCA alorithm, both the nuclear norm penalty parameter and the separation rank are selected by optimizing a normalized Frobenius norm error via grid search.

\begin{figure}
    \centering   
    \begin{subfigure}{\linewidth}
    \centering 
    \includegraphics[width=0.8\linewidth]{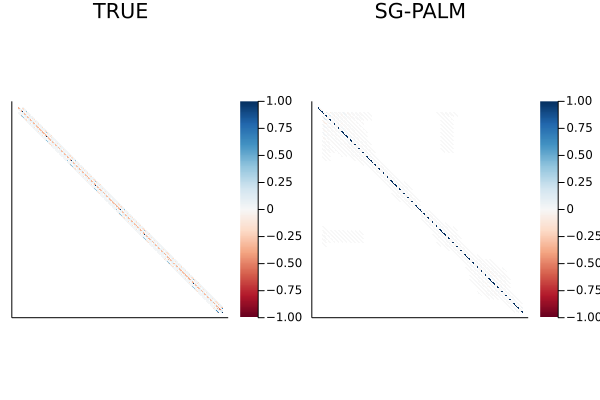}
    \caption{Poisson-AR inverse covariance structure (left) and the estimate obtained by SG-PALM (right).}
    \end{subfigure}
    \begin{subfigure}{\linewidth}
    \centering 
    \includegraphics[width=0.9\linewidth]{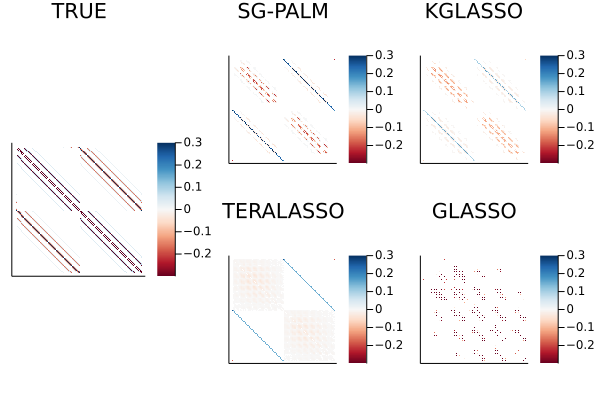}
    \caption{Zoomed-in (middle $128$ rows and columns) Poisson-AR inverse covariance structure (left) and the estimate obtained by SG-PALM, KGlasso, Glasso, TeraLasso (right, clockwise).}
    \end{subfigure}
    \caption{Visualizations of the inverse covariance structures for Poisson-AR(1) dynamics and its estimates. Here, white entries indicate zeros in the inverse covariance matrices. The zoomed-in plots show two temporal blocks (each of size $64 \times 64$) of spatial inverse correlation structures with the diagonal elements removed for clearer visualization. SG-PALM and the associated Sylvester graphical model produce the richest structures.
    }
    \label{fig:poisson_inv_cov_struct_compare}
\end{figure}

\begin{figure}
    \centering   
    \begin{subfigure}{\linewidth}
    \centering 
    \includegraphics[width=0.8\linewidth]{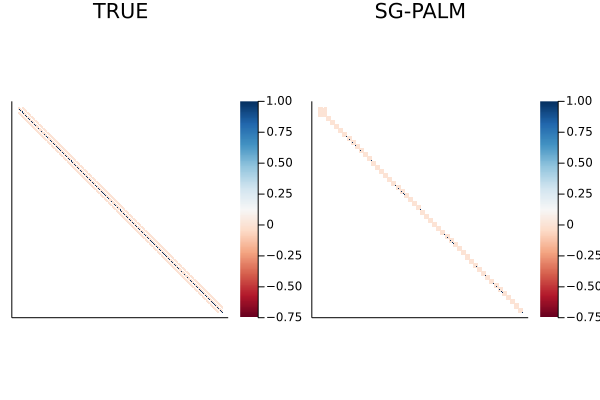}
    \caption{Convection-diffusion inverse covariance structure (left) and the estimate obtained by SG-PALM (right).}
    \end{subfigure}
    \begin{subfigure}{\linewidth}
    \centering 
    \includegraphics[width=0.9\linewidth]{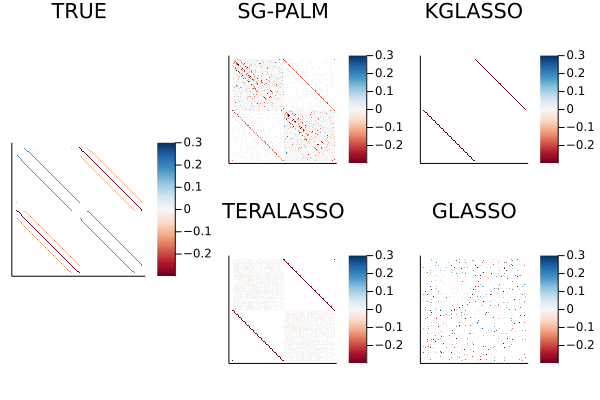}
    \caption{Zoomed-in (middle $128$ rows and columns) convection-diffusion inverse covariance structure (left) and the estimate obtained by SG-PALM, KGlasso, Glasso, TeraLasso (right, clockwise).}
    \end{subfigure}
    \caption{Visualizations of the inverse covariance structures for the Convection-Diffusion dynamics and its estimates. Here, white entries indicate zeros in the inverse covariance matrices. The zoomed-in plots show two temporal blocks (each of size $64 \times 64$) of spatial inverse correlation structures with the diagonal elements removed for clearer visualization. SG-PALM and the associated Sylvester graphical model produce the richest structures.
    }
    \label{fig:convec_diff_inv_cov_struct_compare}
\end{figure}

\begin{figure}
    \centering   
    \begin{subfigure}{\linewidth}
    \centering\includegraphics[width=0.8\linewidth]{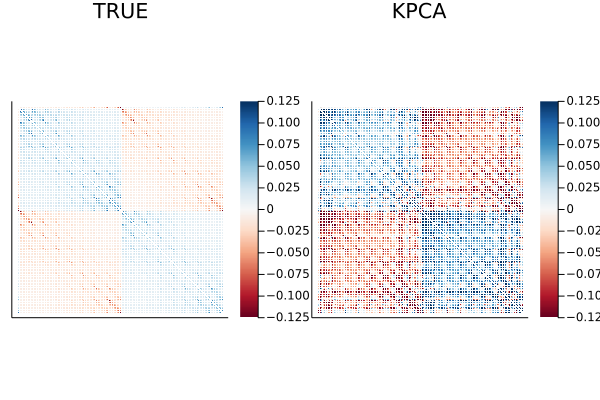}
    \caption{Poisson-AR covariance structure (left) and the estimate obtained by KPCA (right).}
    \end{subfigure}
    \begin{subfigure}{\linewidth}
    \centering\includegraphics[width=0.8\linewidth]{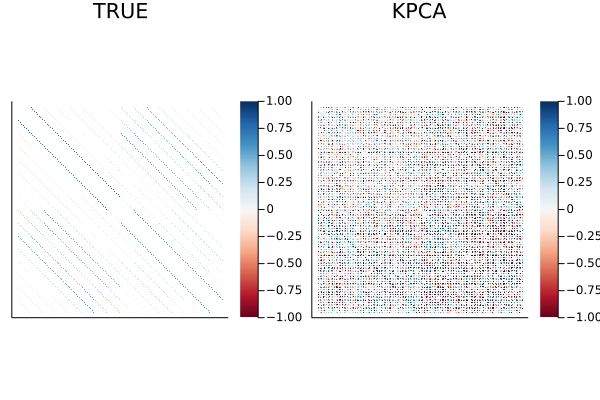}
    \caption{Convection-Diffusion covariance structure (left) and the estimate obtained by KPCA (right).}
    \end{subfigure}
    \caption{Visualizations of the middle $128$ rows and columns of the covariance structures for Poisson-AR(1) and Convection-Diffusion dynamics and their estimates, which show two temporal blocks of spatial correlation structures, each of size $64 \times 64$, with the diagonal elements removed for clearer visualization of the pattern. Here, white entries indicate zeros in the covariance matrices. Since the covariances are not sparse in general, all matrices are thresholded for clearer inspections of patterns.}
    \label{fig:cov_struct_compare}
\end{figure}

A summary of the accuracy of estimation
(normalized Frobenius norm error), as well as support recovery (Mathews Correlation Coefficient~\citep{matthews1975comparison}), are reported in Table~\ref{tab:sythetic_perf}. In Figure~\ref{fig:poisson_inv_cov_struct_compare} and Figure~\ref{fig:convec_diff_inv_cov_struct_compare} we show the true and the estimated inverse covariance matrices obtained for all the methods except KPCA, under both the Poisson-AR (Figure~\ref{fig:poisson_inv_cov_struct_compare}) and the Convection-Diffusion processes (Figure~\ref{fig:convec_diff_inv_cov_struct_compare}). The inverse covariances under both generating processes admit structures with a mix of Kronecker sums and Kronecker products of sparse matrices. In both of the cases, the SG-PALM produces estimates having the lowest error. We believe this is due to the goodness of fit because the Sylvester graphical model imposes a squared KS structure on the precision matrix. Tlasso has comparable performances and achieves the best matrix approximation error under the convection-diffusion generating process. This might be due to the fact that the KP model corresponds to an underlying spatio-temporal autoregressive process. TeraLasso seems to produce the biggest model mismatch as indicated by the MCC scores.
In Figure~\ref{fig:cov_struct_compare}, we compare the true covariance matrices and the estimates obtained by KPCA. Here, although the KPCA model does not match the underlying generating processes, the estimates were able to capture certain blocking patterns that similarly exist in the true covariance.

\begin{table}[!tbh]
\centering
\caption{Comparisons of performances measured by $\log(\|\widehat{\bSigma} - \bSigma\|_F \text{\textbackslash} \|\bSigma\|_F)$ for KPCA as well as $\log(\|\widehat{\bOmega} - \bOmega\|_F \text{\textbackslash} \|\bOmega\|_F)$ and the Mathews Correlation Coefficient (MCC) for SG-PALM, Tlasso, TeraLasso, Glasso. The MCC is a measure of the quality of sparsity recovery considered as a binary classification problem, where $\pm 1$ indicates perfect agreement or disagreement between the truth and the estimation. Here the Frobenius norm errors are included in the first row under each generating type while the MCCs are in the second row. Note that the best performers under each type/criteria are highlighted.}
\label{tab:sythetic_perf}
\begin{tabular}{|c|c||c|c|c|c|c|}
 \multicolumn{7}{c}{} \\
 \hline
 Type & Metric & \textbf{SG-PALM} & \textbf{KGlasso} & \textbf{TeraLasso} & \textbf{Glasso} & \textbf{KPCA}\\
 \cline{1-7} 
 \multirow{2}{*}{P-AR} & Fnorm & $\mathbf{-0.2622}$ & $1.1777$ & $0.6312$ & $0.9775$ & $0.3289$\\
 & MCC & $\mathbf{0.4300}$ & $0.3395$ & $0.2061$ & $0.0560$ & N/A\\
\hline
 \multirow{2}{*}{C-D} & Fnorm & $\mathbf{-0.0420}$ & $1.4919$ & $-0.0208$ & $2.2041$ & $0.0642$\\
 & MCC & $\mathbf{0.2122}$ & $0.1884$ & $0.2018$ & $0.0349$ & N/A\\
 \hline
\end{tabular}
\end{table}

Computational efficiencies of the various covariance/precision estimation algorithms are also vitally important in practice to facilitate real-time tracking of physical systems. Table~\ref{tab:enkf_runtime} shows the runtime of different covariance and inverse covariance estimation algorithms for the synthetic experiments. It shows that by recognizing and exploiting multiway structures in the data, sparse multiway inverse covariance estimation methods, TeraLasso, Tlasso, and SG-PALM significantly reduce the runtime complexity of Glasso that ignores such special multiway structures. Remark that KPCA runs considerably slower than other methods as it involves expensive singular value decomposition of a large-dimensional re-arranged sample covariance matrix of the data.

\begin{table}[tbh!]
\centering
\caption{Average runtime and its standard error (in seconds) of estimating spatio-temporal (inverse) covariance matrices of size $d \times 50$, where $d$ is varying, using various algorithms. Comparisons under various problem sizes (i.e., different $d$ and $N$) are shown. Note the sparse multiway precision models (SG-PALM, KGlasso, TeraLasso) are comparably fast and are all faster than Glasso (for large problems) and KPCA.}
\label{tab:enkf_runtime}
\begin{tabular}{|p{0.3cm}|p{0.3cm}||r|r|r|r|r|}
 \multicolumn{7}{c}{} \\
 \hline
 \multirow{2}{*}{$d$} & \multirow{2}{*}{$N$} & \textbf{Glasso} & \textbf{SG-PALM} & \textbf{TeraLasso} & \textbf{KGlasso} & \textbf{KronPCA}\\
 \cline{3-7} 
 && \textbf{sec} & \textbf{sec} & \textbf{sec} & \textbf{sec} & \textbf{sec} \\
 \hline
 \multirow{3}{*}{$8^2$} & $25$ &
$0.40 (0.20)$ & $0.46 (0.15)$ & $0.15 (0.35)$ & $0.65(0.11)$ & $37.22 (0.20)$ \\
 \cline{3-7}
 & $50$ & 
$0.48 (0.21)$ & $0.47 (0.08)$ & $0.22 (0.50)$ & $0.70 (0.10)$ & $38.22 (0.55)$ \\
 \cline{3-7}
 & $100$ & 
$0.76 (0.05)$ & $0.44 (0.13)$ & $0.26 (0.28)$ & $0.69 (0.30)$ & $39.09 (1.05)$\\
 \cline{1-7}
 \multirow{3}{*}{$16^2$} & $25$ &
$6.43(1.45)$ & $3.37 (1.09)$ & $5.38 (0.58)$ & $5.14 (1.99)$ & $495.47 (2.69)$ \\
 \cline{3-7}
 & $50$ & 
$9.12 (0.98)$ & $3.27 (1.81)$ & $4.62 (1.98)$ & $3.39 (2.00)$ & $516.64 (2.19)$ \\
 \cline{3-7}
 & $100$ & 
 $11.84 (2.01)$ & $4.85 (1.10)$ & $6.71 (0.72)$ & $5.67 (0.57)$ & $498.04 (4.01)$ \\
 \hline
\end{tabular}
\end{table}

\subsection{Application to Predicting Evolution of Solar Active Regions}\label{subsec:ar_pred}
Solar active regions are temporary centers of strong and complex magnetic field on the sun, the principal source of violent eruptions such as solar flares \citep{van2015evolution}. While weak flares of, for example, B-class, have only limited terrestrial effect, strong flares of M- and X-class can produce tremendous amount of electromagnetic radiation, causing disturbance or damage to satellites, power grids, and communication systems. Therefore, it would be great value to be able to predict how active regions evolve before the onset of solar flares. 

Although there are numerous studies that use active region images or physical parameters to predict flare activities \citep{leka2003photospheric,chen2019identifying, jiao2020solar,wang2020predicting,sun2021predicting}, fewer studies have attempted to predict the complicated preflare evolution of active regions without physical modeling \citep{bai2021predicting}. In this section, to demonstrate the application of the tensor graphical models, we use multiwavelength active region observations to predict the evolution of two types of active regions that lead to either a weak (B-class) flare or a strong (M- or X-class) flare~\footnote{A similar and more comprehensive version of this numerical study was included in \citet{wang2021sg}}.

We construct a multiwavelength active region video dataset from the curated dataset generated by \citet{galvez2019machine}. The video data are taken in four wavelengths (94\r{A}, 131\r{A}, 171\r{A}, and 193\r{A}) by the Atmospheric Imaging Assembly \citep[AIA,][]{lemen2011atmospheric} aboard the Solar Dynamics Observatory (SDO) satellite. Each video is a 24-hour image sequence of an active region at 1-hour cadence before a strong (M- or X-class) or a weak (B-class) flare occurs in the region. We spatially interpolate the videos so that each video is represented as a $d_1\times d_2\times d_3 \times d_4$ tensor, where $d_1 = 13$ denotes the number of frames in the video, $d_2 = 50$ denotes the height of the frames after interpolation, $d_3 = 100$ denotes the width of the frames after interpolation, and $d_4=4$ represents the number of different channels/wavelength at which the images are recorded. To prevent information leakage, we chronologically split the active region videos into a training set (year 2011 to 2014) and a test set (year 2011 to 2014). In the training set, there are 186 active region videos that lead to a B-class flare and 48 active region videos that lead to a M/X-class flare. In the test set, the sample sizes are 93 and 24 for the B-class and the M/X-class, respectively.

To perform active region prediction, we first fit the tensor graphical models on the training set to estimate the covariance or prediction matrices for each of the two types of active region videos, and then we use the best linear predictor to predict the last frame from all previous frames for videos in the test set.
The forward linear predictor is constructed in a multi-output least squares regression setting as
\begin{equation}
    \widehat{\mat{y}}_t = -\mat\Omega_{2,2}^{-1}\mat\Omega_{2,1}\mat{y}_{t-1:t-(p-1)}
\end{equation}
when the precision estimate is available. Here, $t = d_1$ for predicting the last frame of a video. For notational convenience, let $p=d_1$ and $q=d_2d_3d_4$, then $\mat{y}_{t-1:t-(p-1)} = \mat{y}_{p-1:1} \in \mathbb{R}^{(p-1)q}$ is the stacked set of pixel values from the previous $p-1$ time instances and $\mat\Omega_{2,1} \in \mathbb{R}^{q \times (p-1)q}$ and $\mat\Omega_{2,2} \in \mathbb{R}^{q \times q}$ are submatrices of the $pq \times pq$ estimated precision matrix:
\begin{equation*}
    \widehat{\mat\Omega} =
    \begin{pmatrix}
    \mat\Omega_{1,1} & \mat\Omega_{1,2} \\
    \mat\Omega_{2,1} & \mat\Omega_{2,2}
    \end{pmatrix}.
\end{equation*}

The tensor graphical models to be evaluated are SG-PALM (Section~\ref{sec:sylvGM}), Tlasso (Section~\ref{sec:teralasso}), and TeraLasso (Section~\ref{sec:KP_precision}), with regularization parameters chosen by optimizating the prediction NRMSE (i.e., the root mean squared error normalized by the difference between maximum and minimum pixels) on the training set over a range of $\lambda$ values guided by the convergence rates presented in Table~\ref{tab:guarantees}. In particular, the SG-PALM estimator was implemented using a regularization parameter $\lambda_{N}=C_1\sqrt{\frac{\min(d_k)\log(d)}{N}}$ for all $k$ with the constant $C_1$ chosen by optimizing the prediction NRMSE on the training set over a range of $\lambda$ values parameterized by $C_1$. The TeraLasso estimator and the Tlasso estimator were implemented using $\lambda_{N,k}=C_2\sqrt{\frac{\log(d)}{N\prod_{i \neq k}d_i}}$ and $\lambda_{N,k}=C_3\sqrt{\frac{\log(d_k)}{Nd}}$ for $k=1,2,3$, respectively, with $C_2, C_3$ optimized in a similar manner.
Also considered is a baseline estimator, referred to as IndLasso, obtained by applying independent and separate $\ell_1$-penalized regressions to each pixel in $\mat{y}_t$.
Each sparse regression in the IndLasso estimator was implemented and tuned independently with regularization parameters chosen in a similar manner.
Note that we did not include Glasso and KPCA (Section~\ref{sec:KP_covar}) in the comparison due to their prohibitively expensive runtime complexity for large multiway data.

\begin{figure}[!tbh]
\centering
\begin{tabular}{@{}c@{}}
    Predicted examples \\
    \rotatebox{90}{\qquad AR B}
    \includegraphics[width=0.95\linewidth]{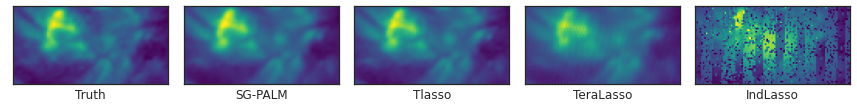}  \\
    \rotatebox{90}{\qquad AR B}
    \includegraphics[width=0.95\linewidth]{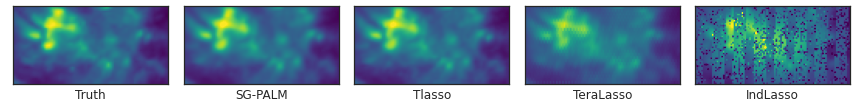} \\
    \rotatebox{90}{\quad AR M/X}
    \includegraphics[width=0.95\linewidth]{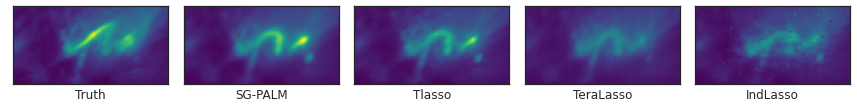} \\
    \rotatebox{90}{\quad AR M/X}
    \includegraphics[width=0.95\linewidth]{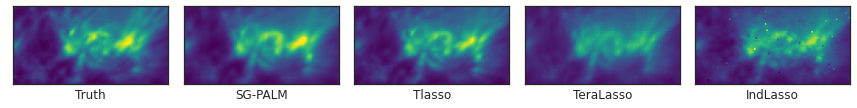}
\end{tabular}
\caption{Examples of one-hour ahead prediction of the first two AIA channels of last frames of $13$-frame videos, leading to B- (first two rows) and M/X-class (last two rows) flares, produced by the SG-PALM, Tlasso, TeraLasso, IndLasso algorithms, 
comparing to the ground truth image (far left column).
The column ordering from left to right is in the order of increasing normalized mean squared error (NMSE) - see Table \ref{tab:ar_pred_nrmse}. Note that in general linear forward predictors tend to underestimate the contrast ratio of the images. The proposed SG-PALM produced the best-quality images in terms of both the spatial structures and contrast ratios. This figure is excerpted from \citet{wang2021sg} (cf. Figure 3 in the reference).}
\label{fig:ar_pred}
\end{figure}

Table~\ref{tab:ar_pred_nrmse} shows the NRMSE over the testing samples for the forecasts based on SG-PALM, Tlasso, TeraLasso, and IndLasso. We observe that SG-PALM outperforms all three other methods, indicated by the lowest NRMSEs on the test set.
Figure~\ref{fig:ar_pred} depicts examples of predicted images, comparing with the ground truth. The SG-PALM estimates produced most realistic image predictions that capture the spatially varying structures and closely approximate the pixel values (i.e., maintaining contrast ratios). The latter is important as the flares are being classified into weak (B-class) and strong (M/X-class) categories based on the brightness of the images, and stronger flares are more likely to lead to catastrophic events, such as those that can lead to damaging terrestrial perturbations and health of astronauts. 

\begin{table}[!tbh]
\centering
\caption{Comparison of the SG-PALM, Tlasso, TeraLasso, IndLasso performances measured by NRMSE in predicting the last frame of $13$-frame video sequences leading to B- and M/X-class solar flares. The NRMSEs are computed by averaging across testing samples and AIA channels for each pixel. SG-PALM achieves the best overall NRMSEs across pixels. B flares are generally easier to predict due to both a larger number of samples in the training set and typically smoother transitions from frame to frame within a video.}
\label{tab:ar_pred_nrmse}
\begin{tabular}{|c||c|c|c|c|}
 \multicolumn{5}{c}{} \\
 \hline
 \multirow{2}{*}{Flare Type} & \textbf{SG-PALM} &
 \textbf{Tlasso} & \textbf{TeraLasso} & \textbf{IndLasso}\\
 \cline{2-5} 
 & NRMSE & NRMSE & NRMSE & NRMSE\\
 \hline
 B & $0.0379$ & $0.0386$ & $0.0579$ & $0.1628$ \\
 M/X & $0.0620$ & $0.0790$ & $0.0913$ & $0.1172$ \\
 \hline
\end{tabular}
\end{table}

\section{Conclusion and discussion}\label{sec:conclusion}
\paragraph{Physical interpretability.} While the Kronecker products expansion used in KPCA captures dense structures in the covariance matrix of data generated from more complex spatio-temporal physical processes, it lacks physical interpretability. In contrast to the case of Sylvester graphical model and Poisson-AR(1) processes, it is not obvious whether the sum of Kronecker products structure corresponds to any true physical models. Recent work in quantum informatics~\citep{chu2021nonlinear} has demonstrated a link between estimation of the density matrix for entangled quantum states and the structured tensor approximation via $\sum_{i=1}^r \mat{A}_i \otimes \mat{B}_i$. Further characterizing and extending these connections to other classes of discretized PDEs is an interesting future direction. Furthermore, in both the Poisson-AR(1) and convection-diffusion examples, a mixed Kronecker sum and Kronecker product structure emerges that can be related to the state covariance of a dynamical system.

\paragraph{Heavy-tailed multiway covariance/precision models.} Most existing work on multiway covariance and inverse covariance models focus on modeling Gaussian variables. It would be interesting to explore whether the pseudo-likelihood framework we adopted for SyGlasso and SG-PALM can be extended to non-Gaussian heavy-tailed models, e.g., using copula's or  elliptically contoured distributions. This could have important practical applications, in particular to solar flare and active region prediction problems presented in Section~\ref{subsec:ar_pred}. The images that characterize the active regions generally include a small number of pixels of extreme high-intensity. These pixels might not be captured by a Gaussian-like distribution. Recently, there have been advances~\citep{wei2017heavytailed,ke2019user} in covariance estimation for heavy-tailed, non-Gaussian vector-variate data. Multiway (inverse) covariance estimation is an open  problem.

\paragraph{Kronecker-structured autoencoders.} Low-rank covariance models have close connections with variational autoencoders (VAEs). \citet{dai2018connections} studied the relationship between (robust) PCA and VAEs. Since the Kronecker product for matrices is a generalization of the outer product for vectors, KPCA can be considered as a generalization of a the low-rank approximation method of PCA. It is thus natural to exploit similar relationships between KPCA and VAEs. In this case VAE may be considered as a nonlinear/non-Gaussian extension to KPCA for low separation rank covariance models. Additionally, recent advances in efficient training of the VAE-type neural network architecture (e.g., using stochastic gradient descent) could improve the computational complexity of KPCA that is currently limited by an expensive singular value decomposition~\citep{tsiligkaridis2013covariance,greenewald2015robust}.

\paragraph{Computational complexity.} 
Estimating the Kronecker product model is generally hard due to: (i) tensor factorization is NP-hard \cite{hillar2013most} when $K \geq 3$; and (ii) the optimization objective is nonconvex. The NP-hardness implies   computational intractability for large scale problems in the worst case. However, this does not necessarily preclude the existence of tractable solutions for most cases.   Specifically, it is believed that the model can be estimated within polynomial time when the signal-to-noise ratio (SNR) is sufficiently high\footnote{For example, the convergence analysis of Tlasso \cite{lyu2019tensor} requires a `good' initialization; a high SNR is required for a random initialization to satisfy the condition.}, but only when it is significantly higher than the information-theoretical threshold for model identifiability \cite{richard2014statistical, hopkins2015tensor}. While such studies have been completed for some important problems e,g, the statistical-computational gap  \cite{chandrasekaran2013computational, zhang2018tensor} and the optimization landscape \cite{ge2020optimization}, an analogous framework for Kronecker covariance remains to be developed.

While Kronecker sum models do not suffer from the nonconvexity of the Kronecker product model, scalable estimation methods that apply to very large  covariance and precision matrices remains computationally challenging. Therefore, it would be worthwhile to develop improvements for the Kronecker sum that mirror the successes in related problems, e.g., introducing more amenable surrogate objectives
\cite{besag1977efficiency, khare2015convex}. 

\paragraph{Software.} A software package called \texttt{TensorGraphicalModels}~\citep{WANG2022100308} has been developed to accompany this review article. Detailed code and instructions for reproducing the numerical results presented in this article (and beyond) are provided in the reference.
The software provides a suite of tools in \texttt{Julia},
which was designed specifically for numerical and scientific computing and has recently become popular for large-scale scientific simulation and modeling, such as those conducted by climate scientists~\citep{perkel2019julia}. 




\clearpage
\bibliographystyle{imsart-nameyear} 
\bibliography{tensor_graphical_models_survey}       

\begin{thebibliography}{106}

\bibitem[\protect\citeauthoryear{Akdemir and Gupta}{2011}]{akdemir2011array}
\begin{barticle}[author]
\bauthor{\bsnm{Akdemir},~\bfnm{Deniz}\binits{D.}} \AND
  \bauthor{\bsnm{Gupta},~\bfnm{Arjun~K}\binits{A.~K.}}
(\byear{2011}).
\btitle{Array Variate Random Variables with Multiway Kronecker Delta Covariance
  Matrix Structure.}
\bjournal{Journal of algebraic statistics}
\bvolume{2}.
\end{barticle}
\endbibitem

\bibitem[\protect\citeauthoryear{Allen and
  Tibshirani}{2010}]{allen2010transposable}
\begin{barticle}[author]
\bauthor{\bsnm{Allen},~\bfnm{Genevera~I}\binits{G.~I.}} \AND
  \bauthor{\bsnm{Tibshirani},~\bfnm{Robert}\binits{R.}}
(\byear{2010}).
\btitle{Transposable regularized covariance models with an application to
  missing data imputation}.
\bjournal{The Annals of Applied Statistics}
\bvolume{4}
\bpages{764}.
\end{barticle}
\endbibitem

\bibitem[\protect\citeauthoryear{Allen and
  Tibshirani}{2012}]{allen2012inference}
\begin{barticle}[author]
\bauthor{\bsnm{Allen},~\bfnm{Genevera~I}\binits{G.~I.}} \AND
  \bauthor{\bsnm{Tibshirani},~\bfnm{Robert}\binits{R.}}
(\byear{2012}).
\btitle{Inference with transposable data: modelling the effects of row and
  column correlations}.
\bjournal{Journal of the Royal Statistical Society: Series B (Statistical
  Methodology)}
\bvolume{74}
\bpages{721--743}.
\end{barticle}
\endbibitem

\bibitem[\protect\citeauthoryear{Bai et~al.}{2021}]{bai2021predicting}
\begin{barticle}[author]
\bauthor{\bsnm{Bai},~\bfnm{Liang}\binits{L.}},
  \bauthor{\bsnm{Bi},~\bfnm{Yi}\binits{Y.}},
  \bauthor{\bsnm{Yang},~\bfnm{Bo}\binits{B.}},
  \bauthor{\bsnm{Hong},~\bfnm{Jun-Chao}\binits{J.-C.}},
  \bauthor{\bsnm{Xu},~\bfnm{Zhe}\binits{Z.}},
  \bauthor{\bsnm{Shang},~\bfnm{Zhen-Hong}\binits{Z.-H.}},
  \bauthor{\bsnm{Liu},~\bfnm{Hui}\binits{H.}},
  \bauthor{\bsnm{Ji},~\bfnm{Hai-Sheng}\binits{H.-S.}} \AND
  \bauthor{\bsnm{Ji},~\bfnm{Kai-Fan}\binits{K.-F.}}
(\byear{2021}).
\btitle{Predicting the evolution of photospheric magnetic field in solar active
  regions using deep learning}.
\bjournal{Research in Astronomy and Astrophysics}
\bvolume{21}
\bpages{113}.
\end{barticle}
\endbibitem

\bibitem[\protect\citeauthoryear{Banerjee, El~Ghaoui and
  d'Aspremont}{2008}]{banerjee2008model}
\begin{barticle}[author]
\bauthor{\bsnm{Banerjee},~\bfnm{Onureena}\binits{O.}},
  \bauthor{\bsnm{El~Ghaoui},~\bfnm{Laurent}\binits{L.}} \AND
  \bauthor{\bsnm{d'Aspremont},~\bfnm{Alexandre}\binits{A.}}
(\byear{2008}).
\btitle{Model selection through sparse maximum likelihood estimation for
  multivariate Gaussian or binary data}.
\bjournal{The Journal of Machine Learning Research}
\bvolume{9}
\bpages{485--516}.
\end{barticle}
\endbibitem

\bibitem[\protect\citeauthoryear{Besag}{1977}]{besag1977efficiency}
\begin{barticle}[author]
\bauthor{\bsnm{Besag},~\bfnm{Julian}\binits{J.}}
(\byear{1977}).
\btitle{Efficiency of pseudolikelihood estimation for simple Gaussian fields}.
\bjournal{Biometrika}
\bpages{616--618}.
\end{barticle}
\endbibitem

\bibitem[\protect\citeauthoryear{Bijma, De~Munck and
  Heethaar}{2005}]{bijma2005spatiotemporal}
\begin{barticle}[author]
\bauthor{\bsnm{Bijma},~\bfnm{Fetsje}\binits{F.}},
  \bauthor{\bsnm{De~Munck},~\bfnm{Jan~C}\binits{J.~C.}} \AND
  \bauthor{\bsnm{Heethaar},~\bfnm{Rob~M}\binits{R.~M.}}
(\byear{2005}).
\btitle{The spatiotemporal MEG covariance matrix modeled as a sum of Kronecker
  products}.
\bjournal{NeuroImage}
\bvolume{27}
\bpages{402--415}.
\end{barticle}
\endbibitem

\bibitem[\protect\citeauthoryear{Box et~al.}{2015}]{box2015time}
\begin{bbook}[author]
\bauthor{\bsnm{Box},~\bfnm{George~EP}\binits{G.~E.}},
  \bauthor{\bsnm{Jenkins},~\bfnm{Gwilym~M}\binits{G.~M.}},
  \bauthor{\bsnm{Reinsel},~\bfnm{Gregory~C}\binits{G.~C.}} \AND
  \bauthor{\bsnm{Ljung},~\bfnm{Greta~M}\binits{G.~M.}}
(\byear{2015}).
\btitle{Time series analysis: forecasting and control}.
\bpublisher{John Wiley \& Sons}.
\end{bbook}
\endbibitem

\bibitem[\protect\citeauthoryear{Cand{\`e}s et~al.}{2011}]{candes2011robust}
\begin{barticle}[author]
\bauthor{\bsnm{Cand{\`e}s},~\bfnm{Emmanuel~J}\binits{E.~J.}},
  \bauthor{\bsnm{Li},~\bfnm{Xiaodong}\binits{X.}},
  \bauthor{\bsnm{Ma},~\bfnm{Yi}\binits{Y.}} \AND
  \bauthor{\bsnm{Wright},~\bfnm{John}\binits{J.}}
(\byear{2011}).
\btitle{Robust principal component analysis?}
\bjournal{Journal of the ACM (JACM)}
\bvolume{58}
\bpages{1--37}.
\end{barticle}
\endbibitem

\bibitem[\protect\citeauthoryear{Chandrasekaran and
  Jordan}{2013}]{chandrasekaran2013computational}
\begin{barticle}[author]
\bauthor{\bsnm{Chandrasekaran},~\bfnm{Venkat}\binits{V.}} \AND
  \bauthor{\bsnm{Jordan},~\bfnm{Michael~I}\binits{M.~I.}}
(\byear{2013}).
\btitle{Computational and statistical tradeoffs via convex relaxation}.
\bjournal{Proceedings of the National Academy of Sciences}
\bvolume{110}
\bpages{E1181--E1190}.
\end{barticle}
\endbibitem

\bibitem[\protect\citeauthoryear{Chandrasekaran
  et~al.}{2011}]{chandrasekaran2011rank}
\begin{barticle}[author]
\bauthor{\bsnm{Chandrasekaran},~\bfnm{Venkat}\binits{V.}},
  \bauthor{\bsnm{Sanghavi},~\bfnm{Sujay}\binits{S.}},
  \bauthor{\bsnm{Parrilo},~\bfnm{Pablo~A}\binits{P.~A.}} \AND
  \bauthor{\bsnm{Willsky},~\bfnm{Alan~S}\binits{A.~S.}}
(\byear{2011}).
\btitle{Rank-sparsity incoherence for matrix decomposition}.
\bjournal{SIAM Journal on Optimization}
\bvolume{21}
\bpages{572--596}.
\end{barticle}
\endbibitem

\bibitem[\protect\citeauthoryear{Chandrasekhar}{1943}]{chandrasekhar1943stochastic}
\begin{barticle}[author]
\bauthor{\bsnm{Chandrasekhar},~\bfnm{Subrahmanyan}\binits{S.}}
(\byear{1943}).
\btitle{Stochastic problems in physics and astronomy}.
\bjournal{Reviews of modern physics}
\bvolume{15}
\bpages{1}.
\end{barticle}
\endbibitem

\bibitem[\protect\citeauthoryear{Chen, Wiesel and Hero}{2011}]{chen2011robust}
\begin{barticle}[author]
\bauthor{\bsnm{Chen},~\bfnm{Yilun}\binits{Y.}},
  \bauthor{\bsnm{Wiesel},~\bfnm{Ami}\binits{A.}} \AND
  \bauthor{\bsnm{Hero},~\bfnm{Alfred~O}\binits{A.~O.}}
(\byear{2011}).
\btitle{Robust shrinkage estimation of high-dimensional covariance matrices}.
\bjournal{IEEE Transactions on Signal Processing}
\bvolume{59}
\bpages{4097--4107}.
\end{barticle}
\endbibitem

\bibitem[\protect\citeauthoryear{Chen et~al.}{2019}]{chen2019identifying}
\begin{barticle}[author]
\bauthor{\bsnm{Chen},~\bfnm{Yang}\binits{Y.}},
  \bauthor{\bsnm{Manchester},~\bfnm{Ward~B}\binits{W.~B.}},
  \bauthor{\bsnm{Hero},~\bfnm{Alfred~O}\binits{A.~O.}},
  \bauthor{\bsnm{Toth},~\bfnm{Gabor}\binits{G.}},
  \bauthor{\bsnm{DuFumier},~\bfnm{Benoit}\binits{B.}},
  \bauthor{\bsnm{Zhou},~\bfnm{Tian}\binits{T.}},
  \bauthor{\bsnm{Wang},~\bfnm{Xiantong}\binits{X.}},
  \bauthor{\bsnm{Zhu},~\bfnm{Haonan}\binits{H.}},
  \bauthor{\bsnm{Sun},~\bfnm{Zeyu}\binits{Z.}} \AND
  \bauthor{\bsnm{Gombosi},~\bfnm{Tamas~I}\binits{T.~I.}}
(\byear{2019}).
\btitle{Identifying solar flare precursors using time series of SDO/HMI images
  and SHARP parameters}.
\bjournal{Space Weather}
\bvolume{17}
\bpages{1404--1426}.
\end{barticle}
\endbibitem

\bibitem[\protect\citeauthoryear{Chu and Lin}{2021}]{chu2021nonlinear}
\begin{barticle}[author]
\bauthor{\bsnm{Chu},~\bfnm{Moody~T}\binits{M.~T.}} \AND
  \bauthor{\bsnm{Lin},~\bfnm{Matthew~M}\binits{M.~M.}}
(\byear{2021}).
\btitle{Nonlinear Power-Like and SVD-Like Iterative Schemes with Applications
  to Entangled Bipartite Rank-1 Approximation}.
\bjournal{SIAM Journal on Scientific Computing}
\bvolume{0}
\bpages{S448--S474}.
\end{barticle}
\endbibitem

\bibitem[\protect\citeauthoryear{Cressie}{2015}]{cressie2015statistics}
\begin{bbook}[author]
\bauthor{\bsnm{Cressie},~\bfnm{Noel}\binits{N.}}
(\byear{2015}).
\btitle{Statistics for spatial data}.
\bpublisher{John Wiley \& Sons}.
\end{bbook}
\endbibitem

\bibitem[\protect\citeauthoryear{Dai et~al.}{2018}]{dai2018connections}
\begin{barticle}[author]
\bauthor{\bsnm{Dai},~\bfnm{Bin}\binits{B.}},
  \bauthor{\bsnm{Wang},~\bfnm{Yu}\binits{Y.}},
  \bauthor{\bsnm{Aston},~\bfnm{John}\binits{J.}},
  \bauthor{\bsnm{Hua},~\bfnm{Gang}\binits{G.}} \AND
  \bauthor{\bsnm{Wipf},~\bfnm{David}\binits{D.}}
(\byear{2018}).
\btitle{Connections with robust PCA and the role of emergent sparsity in
  variational autoencoder models}.
\bjournal{The Journal of Machine Learning Research}
\bvolume{19}
\bpages{1573--1614}.
\end{barticle}
\endbibitem

\bibitem[\protect\citeauthoryear{Dantas, Cohen and
  Gribonval}{2019}]{dantas2019learning}
\begin{binproceedings}[author]
\bauthor{\bsnm{Dantas},~\bfnm{C{\'a}ssio~F}\binits{C.~F.}},
  \bauthor{\bsnm{Cohen},~\bfnm{J{\'e}r{\'e}my~E}\binits{J.~E.}} \AND
  \bauthor{\bsnm{Gribonval},~\bfnm{R{\'e}mi}\binits{R.}}
(\byear{2019}).
\btitle{Learning tensor-structured dictionaries with application to
  hyperspectral image denoising}.
In \bbooktitle{2019 27th European Signal Processing Conference (EUSIPCO)}
\bpages{1--5}.
\bpublisher{IEEE}.
\end{binproceedings}
\endbibitem

\bibitem[\protect\citeauthoryear{Dawid}{1981}]{dawid1981some}
\begin{barticle}[author]
\bauthor{\bsnm{Dawid},~\bfnm{A~Philip}\binits{A.~P.}}
(\byear{1981}).
\btitle{Some matrix-variate distribution theory: notational considerations and
  a Bayesian application}.
\bjournal{Biometrika}
\bvolume{68}
\bpages{265--274}.
\end{barticle}
\endbibitem

\bibitem[\protect\citeauthoryear{Deng, Tang and Qu}{2021}]{deng2021correlation}
\begin{barticle}[author]
\bauthor{\bsnm{Deng},~\bfnm{Yujia}\binits{Y.}},
  \bauthor{\bsnm{Tang},~\bfnm{Xiwei}\binits{X.}} \AND
  \bauthor{\bsnm{Qu},~\bfnm{Annie}\binits{A.}}
(\byear{2021}).
\btitle{Correlation Tensor Decomposition and Its Application in Spatial Imaging
  Data}.
\bjournal{Journal of the American Statistical Association}
\bvolume{just-accepted}
\bpages{1--35}.
\end{barticle}
\endbibitem

\bibitem[\protect\citeauthoryear{Du et~al.}{2017}]{du2017cbinderdb}
\begin{barticle}[author]
\bauthor{\bsnm{Du},~\bfnm{Jiewen}\binits{J.}},
  \bauthor{\bsnm{Yan},~\bfnm{Xin}\binits{X.}},
  \bauthor{\bsnm{Liu},~\bfnm{Zhihong}\binits{Z.}},
  \bauthor{\bsnm{Cui},~\bfnm{Lu}\binits{L.}},
  \bauthor{\bsnm{Ding},~\bfnm{Peng}\binits{P.}},
  \bauthor{\bsnm{Tan},~\bfnm{Xiaoqing}\binits{X.}},
  \bauthor{\bsnm{Li},~\bfnm{Xiuming}\binits{X.}},
  \bauthor{\bsnm{Zhou},~\bfnm{Huihao}\binits{H.}},
  \bauthor{\bsnm{Gu},~\bfnm{Qiong}\binits{Q.}} \AND
  \bauthor{\bsnm{Xu},~\bfnm{Jun}\binits{J.}}
(\byear{2017}).
\btitle{cBinderDB: a covalent binding agent database}.
\bjournal{Bioinformatics}
\bvolume{33}
\bpages{1258--1260}.
\end{barticle}
\endbibitem

\bibitem[\protect\citeauthoryear{Dutilleul}{1999}]{dutilleul1999mle}
\begin{barticle}[author]
\bauthor{\bsnm{Dutilleul},~\bfnm{Pierre}\binits{P.}}
(\byear{1999}).
\btitle{The MLE algorithm for the matrix normal distribution}.
\bjournal{Journal of statistical computation and simulation}
\bvolume{64}
\bpages{105--123}.
\end{barticle}
\endbibitem

\bibitem[\protect\citeauthoryear{Evensen}{1994}]{evensen1994sequential}
\begin{barticle}[author]
\bauthor{\bsnm{Evensen},~\bfnm{Geir}\binits{G.}}
(\byear{1994}).
\btitle{Sequential data assimilation with a nonlinear quasi-geostrophic model
  using Monte Carlo methods to forecast error statistics}.
\bjournal{Journal of Geophysical Research: Oceans}
\bvolume{99}
\bpages{10143--10162}.
\end{barticle}
\endbibitem

\bibitem[\protect\citeauthoryear{Friedman, Hastie and
  Tibshirani}{2008}]{friedman2008sparse}
\begin{barticle}[author]
\bauthor{\bsnm{Friedman},~\bfnm{Jerome}\binits{J.}},
  \bauthor{\bsnm{Hastie},~\bfnm{Trevor}\binits{T.}} \AND
  \bauthor{\bsnm{Tibshirani},~\bfnm{Robert}\binits{R.}}
(\byear{2008}).
\btitle{Sparse inverse covariance estimation with the graphical lasso}.
\bjournal{Biostatistics}
\bvolume{9}
\bpages{432--441}.
\end{barticle}
\endbibitem

\bibitem[\protect\citeauthoryear{Fu and Huang}{2008}]{fu2008image}
\begin{barticle}[author]
\bauthor{\bsnm{Fu},~\bfnm{Yun}\binits{Y.}} \AND
  \bauthor{\bsnm{Huang},~\bfnm{Thomas~S}\binits{T.~S.}}
(\byear{2008}).
\btitle{Image classification using correlation tensor analysis}.
\bjournal{IEEE Transactions on Image Processing}
\bvolume{17}
\bpages{226--234}.
\end{barticle}
\endbibitem

\bibitem[\protect\citeauthoryear{Galecki}{1994}]{galecki1994general}
\begin{barticle}[author]
\bauthor{\bsnm{Galecki},~\bfnm{Andrzej~T}\binits{A.~T.}}
(\byear{1994}).
\btitle{General class of covariance structures for two or more repeated factors
  in longitudinal data analysis}.
\bjournal{Communications in Statistics-Theory and Methods}
\bvolume{23}
\bpages{3105--3119}.
\end{barticle}
\endbibitem

\bibitem[\protect\citeauthoryear{Galvez et~al.}{2019}]{galvez2019machine}
\begin{barticle}[author]
\bauthor{\bsnm{Galvez},~\bfnm{Richard}\binits{R.}},
  \bauthor{\bsnm{Fouhey},~\bfnm{David~F}\binits{D.~F.}},
  \bauthor{\bsnm{Jin},~\bfnm{Meng}\binits{M.}},
  \bauthor{\bsnm{Szenicer},~\bfnm{Alexandre}\binits{A.}},
  \bauthor{\bsnm{Mu{\~n}oz-Jaramillo},~\bfnm{Andr{\'e}s}\binits{A.}},
  \bauthor{\bsnm{Cheung},~\bfnm{Mark~CM}\binits{M.~C.}},
  \bauthor{\bsnm{Wright},~\bfnm{Paul~J}\binits{P.~J.}},
  \bauthor{\bsnm{Bobra},~\bfnm{Monica~G}\binits{M.~G.}},
  \bauthor{\bsnm{Liu},~\bfnm{Yang}\binits{Y.}},
  \bauthor{\bsnm{Mason},~\bfnm{James}\binits{J.}} \betal{et~al.}
(\byear{2019}).
\btitle{A machine-learning data set prepared from the NASA solar dynamics
  observatory mission}.
\bjournal{The Astrophysical Journal Supplement Series}
\bvolume{242}
\bpages{7}.
\end{barticle}
\endbibitem

\bibitem[\protect\citeauthoryear{Ge and Ma}{2020}]{ge2020optimization}
\begin{barticle}[author]
\bauthor{\bsnm{Ge},~\bfnm{Rong}\binits{R.}} \AND
  \bauthor{\bsnm{Ma},~\bfnm{Tengyu}\binits{T.}}
(\byear{2020}).
\btitle{On the optimization landscape of tensor decompositions}.
\bjournal{Mathematical Programming}
\bpages{1--47}.
\end{barticle}
\endbibitem

\bibitem[\protect\citeauthoryear{Grasedyck}{2004}]{grasedyck2004existence}
\begin{barticle}[author]
\bauthor{\bsnm{Grasedyck},~\bfnm{Lars}\binits{L.}}
(\byear{2004}).
\btitle{Existence and computation of low Kronecker-rank approximations for
  large linear systems of tensor product structure}.
\bjournal{Computing}
\bvolume{72}
\bpages{247--265}.
\end{barticle}
\endbibitem

\bibitem[\protect\citeauthoryear{Greenewald and
  Hero}{2014}]{greenewald2014kronecker}
\begin{binproceedings}[author]
\bauthor{\bsnm{Greenewald},~\bfnm{Kristjan~H}\binits{K.~H.}} \AND
  \bauthor{\bsnm{Hero},~\bfnm{Alfred~O}\binits{A.~O.}}
(\byear{2014}).
\btitle{Kronecker PCA based spatio-temporal modeling of video for dismount
  classification}.
In \bbooktitle{Algorithms for Synthetic Aperture Radar Imagery XXI}
\bvolume{9093}
\bpages{90930V}.
\bpublisher{International Society for Optics and Photonics}.
\end{binproceedings}
\endbibitem

\bibitem[\protect\citeauthoryear{Greenewald and
  Hero}{2015}]{greenewald2015robust}
\begin{barticle}[author]
\bauthor{\bsnm{Greenewald},~\bfnm{Kristjan}\binits{K.}} \AND
  \bauthor{\bsnm{Hero},~\bfnm{Alfred~O}\binits{A.~O.}}
(\byear{2015}).
\btitle{Robust kronecker product PCA for spatio-temporal covariance
  estimation}.
\bjournal{IEEE Transactions on Signal Processing}
\bvolume{63}
\bpages{6368--6378}.
\end{barticle}
\endbibitem

\bibitem[\protect\citeauthoryear{Greenewald, Zelnio and
  Hero}{2016}]{greenewald2016robust}
\begin{barticle}[author]
\bauthor{\bsnm{Greenewald},~\bfnm{Kristjan}\binits{K.}},
  \bauthor{\bsnm{Zelnio},~\bfnm{Edmund}\binits{E.}} \AND
  \bauthor{\bsnm{Hero},~\bfnm{Alfred~Hero}\binits{A.~H.}}
(\byear{2016}).
\btitle{Robust SAR STAP via Kronecker decomposition}.
\bjournal{IEEE Transactions on Aerospace and Electronic Systems}
\bvolume{52}
\bpages{2612--2625}.
\end{barticle}
\endbibitem

\bibitem[\protect\citeauthoryear{Greenewald, Zhou and
  Hero}{2019}]{greenewald2019tensor}
\begin{barticle}[author]
\bauthor{\bsnm{Greenewald},~\bfnm{Kristjan}\binits{K.}},
  \bauthor{\bsnm{Zhou},~\bfnm{Shuheng}\binits{S.}} \AND
  \bauthor{\bsnm{Hero},~\bfnm{Alfred}\binits{A.}}
(\byear{2019}).
\btitle{Tensor graphical lasso (TeraLasso)}.
\bjournal{Journal of the Royal Statistical Society: Series B (Statistical
  Methodology)}
\bvolume{81}
\bpages{901--931}.
\end{barticle}
\endbibitem

\bibitem[\protect\citeauthoryear{Hao et~al.}{2021}]{hao2021sparse}
\begin{barticle}[author]
\bauthor{\bsnm{Hao},~\bfnm{Botao}\binits{B.}},
  \bauthor{\bsnm{Wang},~\bfnm{Boxiang}\binits{B.}},
  \bauthor{\bsnm{Wang},~\bfnm{Pengyuan}\binits{P.}},
  \bauthor{\bsnm{Zhang},~\bfnm{Jingfei}\binits{J.}},
  \bauthor{\bsnm{Yang},~\bfnm{Jian}\binits{J.}} \AND
  \bauthor{\bsnm{Sun},~\bfnm{Will~Wei}\binits{W.~W.}}
(\byear{2021}).
\btitle{Sparse tensor additive regression}.
\bjournal{Journal of Machine Learning Research}
\bvolume{22}
\bpages{1--43}.
\end{barticle}
\endbibitem

\bibitem[\protect\citeauthoryear{Hillar and Lim}{2013}]{hillar2013most}
\begin{barticle}[author]
\bauthor{\bsnm{Hillar},~\bfnm{Christopher~J}\binits{C.~J.}} \AND
  \bauthor{\bsnm{Lim},~\bfnm{Lek-Heng}\binits{L.-H.}}
(\byear{2013}).
\btitle{Most tensor problems are NP-hard}.
\bjournal{Journal of the ACM (JACM)}
\bvolume{60}
\bpages{1--39}.
\end{barticle}
\endbibitem

\bibitem[\protect\citeauthoryear{Hoff}{2011}]{hoff2011separable}
\begin{barticle}[author]
\bauthor{\bsnm{Hoff},~\bfnm{Peter~D}\binits{P.~D.}}
(\byear{2011}).
\btitle{Separable covariance arrays via the Tucker product, with applications
  to multivariate relational data}.
\bjournal{Bayesian Analysis}
\bvolume{6}
\bpages{179--196}.
\end{barticle}
\endbibitem

\bibitem[\protect\citeauthoryear{Hopkins, Shi and
  Steurer}{2015}]{hopkins2015tensor}
\begin{binproceedings}[author]
\bauthor{\bsnm{Hopkins},~\bfnm{Samuel~B}\binits{S.~B.}},
  \bauthor{\bsnm{Shi},~\bfnm{Jonathan}\binits{J.}} \AND
  \bauthor{\bsnm{Steurer},~\bfnm{David}\binits{D.}}
(\byear{2015}).
\btitle{Tensor principal component analysis via sum-of-square proofs}.
In \bbooktitle{Conference on Learning Theory}
\bpages{956--1006}.
\bpublisher{PMLR}.
\end{binproceedings}
\endbibitem

\bibitem[\protect\citeauthoryear{Hori et~al.}{2016}]{hori2016multi}
\begin{barticle}[author]
\bauthor{\bsnm{Hori},~\bfnm{Tomoaki}\binits{T.}},
  \bauthor{\bsnm{Montcho},~\bfnm{David}\binits{D.}},
  \bauthor{\bsnm{Agbangla},~\bfnm{Clement}\binits{C.}},
  \bauthor{\bsnm{Ebana},~\bfnm{Kaworu}\binits{K.}},
  \bauthor{\bsnm{Futakuchi},~\bfnm{Koichi}\binits{K.}} \AND
  \bauthor{\bsnm{Iwata},~\bfnm{Hiroyoshi}\binits{H.}}
(\byear{2016}).
\btitle{Multi-task Gaussian process for imputing missing data in multi-trait
  and multi-environment trials}.
\bjournal{Theoretical and Applied Genetics}
\bvolume{129}
\bpages{2101--2115}.
\end{barticle}
\endbibitem

\bibitem[\protect\citeauthoryear{Hou, Lawrence and
  Hero}{2021}]{hou2021penalized}
\begin{barticle}[author]
\bauthor{\bsnm{Hou},~\bfnm{Elizabeth}\binits{E.}},
  \bauthor{\bsnm{Lawrence},~\bfnm{Earl}\binits{E.}} \AND
  \bauthor{\bsnm{Hero},~\bfnm{Alfred~O}\binits{A.~O.}}
(\byear{2021}).
\btitle{Penalized ensemble Kalman filters for high dimensional non-linear
  systems}.
\bjournal{PloS one}
\bvolume{16}
\bpages{e0248046}.
\end{barticle}
\endbibitem

\bibitem[\protect\citeauthoryear{Jiang, Bigot and
  Maabout}{2021}]{jiang2021online}
\begin{barticle}[author]
\bauthor{\bsnm{Jiang},~\bfnm{Yiye}\binits{Y.}},
  \bauthor{\bsnm{Bigot},~\bfnm{J{\'e}r{\'e}mie}\binits{J.}} \AND
  \bauthor{\bsnm{Maabout},~\bfnm{Sofian}\binits{S.}}
(\byear{2021}).
\btitle{Online Graph Topology Learning from Matrix-valued Time Series}.
\bjournal{arXiv preprint arXiv:2107.08020}.
\end{barticle}
\endbibitem

\bibitem[\protect\citeauthoryear{Jiao et~al.}{2020}]{jiao2020solar}
\begin{barticle}[author]
\bauthor{\bsnm{Jiao},~\bfnm{Zhenbang}\binits{Z.}},
  \bauthor{\bsnm{Sun},~\bfnm{Hu}\binits{H.}},
  \bauthor{\bsnm{Wang},~\bfnm{Xiantong}\binits{X.}},
  \bauthor{\bsnm{Manchester},~\bfnm{Ward}\binits{W.}},
  \bauthor{\bsnm{Gombosi},~\bfnm{Tamas}\binits{T.}},
  \bauthor{\bsnm{Hero},~\bfnm{Alfred}\binits{A.}} \AND
  \bauthor{\bsnm{Chen},~\bfnm{Yang}\binits{Y.}}
(\byear{2020}).
\btitle{Solar flare intensity prediction with machine learning models}.
\bjournal{Space Weather}
\bvolume{18}
\bpages{e2020SW002440}.
\end{barticle}
\endbibitem

\bibitem[\protect\citeauthoryear{Jolliffe}{1986}]{jolliffe1986principal}
\begin{bincollection}[author]
\bauthor{\bsnm{Jolliffe},~\bfnm{Ian~T}\binits{I.~T.}}
(\byear{1986}).
\btitle{Principal components in regression analysis}.
In \bbooktitle{Principal component analysis}
\bpages{129--155}.
\bpublisher{Springer}.
\end{bincollection}
\endbibitem

\bibitem[\protect\citeauthoryear{Kalaitzis
  et~al.}{2013}]{kalaitzis2013bigraphical}
\begin{binproceedings}[author]
\bauthor{\bsnm{Kalaitzis},~\bfnm{Alfredo}\binits{A.}},
  \bauthor{\bsnm{Lafferty},~\bfnm{John}\binits{J.}},
  \bauthor{\bsnm{Lawrence},~\bfnm{Neil~D}\binits{N.~D.}} \AND
  \bauthor{\bsnm{Zhou},~\bfnm{Shuheng}\binits{S.}}
(\byear{2013}).
\btitle{The bigraphical lasso}.
In \bbooktitle{International Conference on Machine Learning}
\bpages{1229--1237}.
\bpublisher{PMLR}.
\end{binproceedings}
\endbibitem

\bibitem[\protect\citeauthoryear{Ke et~al.}{2019}]{ke2019user}
\begin{barticle}[author]
\bauthor{\bsnm{Ke},~\bfnm{Yuan}\binits{Y.}},
  \bauthor{\bsnm{Minsker},~\bfnm{Stanislav}\binits{S.}},
  \bauthor{\bsnm{Ren},~\bfnm{Zhao}\binits{Z.}},
  \bauthor{\bsnm{Sun},~\bfnm{Qiang}\binits{Q.}} \AND
  \bauthor{\bsnm{Zhou},~\bfnm{Wen-Xin}\binits{W.-X.}}
(\byear{2019}).
\btitle{User-friendly covariance estimation for heavy-tailed distributions}.
\bjournal{Statistical Science}
\bvolume{34}
\bpages{454--471}.
\end{barticle}
\endbibitem

\bibitem[\protect\citeauthoryear{Khare, Oh and
  Rajaratnam}{2015}]{khare2015convex}
\begin{barticle}[author]
\bauthor{\bsnm{Khare},~\bfnm{Kshitij}\binits{K.}},
  \bauthor{\bsnm{Oh},~\bfnm{Sang-Yun}\binits{S.-Y.}} \AND
  \bauthor{\bsnm{Rajaratnam},~\bfnm{Bala}\binits{B.}}
(\byear{2015}).
\btitle{A convex pseudolikelihood framework for high dimensional partial
  correlation estimation with convergence guarantees}.
\bjournal{Journal of the Royal Statistical Society: Series B: Statistical
  Methodology}
\bpages{803--825}.
\end{barticle}
\endbibitem

\bibitem[\protect\citeauthoryear{Kolda and Bader}{2009}]{kolda2009tensor}
\begin{barticle}[author]
\bauthor{\bsnm{Kolda},~\bfnm{Tamara~G}\binits{T.~G.}} \AND
  \bauthor{\bsnm{Bader},~\bfnm{Brett~W}\binits{B.~W.}}
(\byear{2009}).
\btitle{Tensor decompositions and applications}.
\bjournal{SIAM review}
\bvolume{51}
\bpages{455--500}.
\end{barticle}
\endbibitem

\bibitem[\protect\citeauthoryear{Kressner and
  Tobler}{2010}]{kressner2010krylov}
\begin{barticle}[author]
\bauthor{\bsnm{Kressner},~\bfnm{Daniel}\binits{D.}} \AND
  \bauthor{\bsnm{Tobler},~\bfnm{Christine}\binits{C.}}
(\byear{2010}).
\btitle{Krylov subspace methods for linear systems with tensor product
  structure}.
\bjournal{SIAM journal on matrix analysis and applications}
\bvolume{31}
\bpages{1688--1714}.
\end{barticle}
\endbibitem

\bibitem[\protect\citeauthoryear{Landsberg}{2011}]{landsberg2011tensors}
\begin{bbook}[author]
\bauthor{\bsnm{Landsberg},~\bfnm{JM}\binits{J.}}
(\byear{2011}).
\btitle{Tensors: Geometry and Applications: Geometry and Applications}
\bvolume{128}.
\bpublisher{American Mathematical Soc.}
\end{bbook}
\endbibitem

\bibitem[\protect\citeauthoryear{Ledoit and Wolf}{2020}]{ledoit2020power}
\begin{barticle}[author]
\bauthor{\bsnm{Ledoit},~\bfnm{Olivier}\binits{O.}} \AND
  \bauthor{\bsnm{Wolf},~\bfnm{Michael}\binits{M.}}
(\byear{2020}).
\btitle{The power of (non-) linear shrinking: A review and guide to covariance
  matrix estimation}.
\bjournal{Journal of Financial Econometrics}.
\end{barticle}
\endbibitem

\bibitem[\protect\citeauthoryear{Leka and Barnes}{2003}]{leka2003photospheric}
\begin{barticle}[author]
\bauthor{\bsnm{Leka},~\bfnm{KD}\binits{K.}} \AND
  \bauthor{\bsnm{Barnes},~\bfnm{G}\binits{G.}}
(\byear{2003}).
\btitle{Photospheric magnetic field properties of flaring versus flare-quiet
  active regions. II. Discriminant analysis}.
\bjournal{The Astrophysical Journal}
\bvolume{595}
\bpages{1296}.
\end{barticle}
\endbibitem

\bibitem[\protect\citeauthoryear{Lemen et~al.}{2011}]{lemen2011atmospheric}
\begin{bincollection}[author]
\bauthor{\bsnm{Lemen},~\bfnm{James~R}\binits{J.~R.}},
  \bauthor{\bsnm{Akin},~\bfnm{David~J}\binits{D.~J.}},
  \bauthor{\bsnm{Boerner},~\bfnm{Paul~F}\binits{P.~F.}},
  \bauthor{\bsnm{Chou},~\bfnm{Catherine}\binits{C.}},
  \bauthor{\bsnm{Drake},~\bfnm{Jerry~F}\binits{J.~F.}},
  \bauthor{\bsnm{Duncan},~\bfnm{Dexter~W}\binits{D.~W.}},
  \bauthor{\bsnm{Edwards},~\bfnm{Christopher~G}\binits{C.~G.}},
  \bauthor{\bsnm{Friedlaender},~\bfnm{Frank~M}\binits{F.~M.}},
  \bauthor{\bsnm{Heyman},~\bfnm{Gary~F}\binits{G.~F.}},
  \bauthor{\bsnm{Hurlburt},~\bfnm{Neal~E}\binits{N.~E.}} \betal{et~al.}
(\byear{2011}).
\btitle{The atmospheric imaging assembly (AIA) on the solar dynamics
  observatory (SDO)}.
In \bbooktitle{The solar dynamics observatory}
\bpages{17--40}.
\bpublisher{Springer}.
\end{bincollection}
\endbibitem

\bibitem[\protect\citeauthoryear{Leng and Pan}{2018}]{leng2018covariance}
\begin{barticle}[author]
\bauthor{\bsnm{Leng},~\bfnm{Chenlei}\binits{C.}} \AND
  \bauthor{\bsnm{Pan},~\bfnm{Guangming}\binits{G.}}
(\byear{2018}).
\btitle{Covariance estimation via sparse Kronecker structures}.
\bjournal{Bernoulli}
\bvolume{24}
\bpages{3833--3863}.
\end{barticle}
\endbibitem

\bibitem[\protect\citeauthoryear{Li and Zhang}{2017}]{li2017parsimonious}
\begin{barticle}[author]
\bauthor{\bsnm{Li},~\bfnm{Lexin}\binits{L.}} \AND
  \bauthor{\bsnm{Zhang},~\bfnm{Xin}\binits{X.}}
(\byear{2017}).
\btitle{Parsimonious tensor response regression}.
\bjournal{Journal of the American Statistical Association}
\bvolume{112}
\bpages{1131--1146}.
\end{barticle}
\endbibitem

\bibitem[\protect\citeauthoryear{Li et~al.}{2008}]{li2008three}
\begin{barticle}[author]
\bauthor{\bsnm{Li},~\bfnm{Zhijin}\binits{Z.}},
  \bauthor{\bsnm{Chao},~\bfnm{Yi}\binits{Y.}},
  \bauthor{\bsnm{McWilliams},~\bfnm{James~C}\binits{J.~C.}} \AND
  \bauthor{\bsnm{Ide},~\bfnm{Kayo}\binits{K.}}
(\byear{2008}).
\btitle{A three-dimensional variational data assimilation scheme for the
  regional ocean modeling system}.
\bjournal{Journal of Atmospheric and Oceanic Technology}
\bvolume{25}
\bpages{2074--2090}.
\end{barticle}
\endbibitem

\bibitem[\protect\citeauthoryear{Li et~al.}{2022}]{li2022scalable}
\begin{barticle}[author]
\bauthor{\bsnm{Li},~\bfnm{Sijia}\binits{S.}},
  \bauthor{\bsnm{L{\'o}pez-Garc{\'\i}a},~\bfnm{Mart{\'\i}n}\binits{M.}},
  \bauthor{\bsnm{Lawrence},~\bfnm{Neil~D}\binits{N.~D.}} \AND
  \bauthor{\bsnm{Cutillo},~\bfnm{Luisa}\binits{L.}}
(\byear{2022}).
\btitle{Scalable Bigraphical Lasso: Two-way Sparse Network Inference for Count
  Data}.
\bjournal{arXiv preprint arXiv:2203.07912}.
\end{barticle}
\endbibitem

\bibitem[\protect\citeauthoryear{Lindgren, Rue and
  Lindstr{\"o}m}{2011}]{lindgren2011explicit}
\begin{barticle}[author]
\bauthor{\bsnm{Lindgren},~\bfnm{Finn}\binits{F.}},
  \bauthor{\bsnm{Rue},~\bfnm{H{\aa}vard}\binits{H.}} \AND
  \bauthor{\bsnm{Lindstr{\"o}m},~\bfnm{Johan}\binits{J.}}
(\byear{2011}).
\btitle{An explicit link between Gaussian fields and Gaussian Markov random
  fields: the stochastic partial differential equation approach}.
\bjournal{Journal of the Royal Statistical Society: Series B (Statistical
  Methodology)}
\bvolume{73}
\bpages{423--498}.
\end{barticle}
\endbibitem

\bibitem[\protect\citeauthoryear{Liu et~al.}{2019}]{liu2019regional}
\begin{barticle}[author]
\bauthor{\bsnm{Liu},~\bfnm{Gang}\binits{G.}},
  \bauthor{\bsnm{Tan},~\bfnm{Xiaoqing}\binits{X.}},
  \bauthor{\bsnm{Dang},~\bfnm{Chao}\binits{C.}},
  \bauthor{\bsnm{Tan},~\bfnm{Shuangquan}\binits{S.}},
  \bauthor{\bsnm{Xing},~\bfnm{Shihui}\binits{S.}},
  \bauthor{\bsnm{Huang},~\bfnm{Nianwei}\binits{N.}},
  \bauthor{\bsnm{Peng},~\bfnm{Kangqiang}\binits{K.}},
  \bauthor{\bsnm{Xie},~\bfnm{Chuanmiao}\binits{C.}},
  \bauthor{\bsnm{Tang},~\bfnm{Xiaoying}\binits{X.}} \AND
  \bauthor{\bsnm{Zeng},~\bfnm{Jinsheng}\binits{J.}}
(\byear{2019}).
\btitle{Regional shape abnormalities in thalamus and verbal memory impairment
  after subcortical infarction}.
\bjournal{Neurorehabilitation and Neural Repair}
\bvolume{33}
\bpages{476--485}.
\end{barticle}
\endbibitem

\bibitem[\protect\citeauthoryear{Llosa-Vite and
  Maitra}{2022}]{llosa2020reduced}
\begin{barticle}[author]
\bauthor{\bsnm{Llosa-Vite},~\bfnm{Carlos}\binits{C.}} \AND
  \bauthor{\bsnm{Maitra},~\bfnm{Ranjan}\binits{R.}}
(\byear{2022}).
\btitle{Reduced-Rank Tensor-on-Tensor Regression and Tensor-variate Analysis of
  Variance}.
\bjournal{IEEE Transactions on Pattern Analysis and Machine Intelligence}
\bpages{1-1}.
\bdoi{10.1109/TPAMI.2022.3164836}
\end{barticle}
\endbibitem

\bibitem[\protect\citeauthoryear{Lyu et~al.}{2019}]{lyu2019tensor}
\begin{barticle}[author]
\bauthor{\bsnm{Lyu},~\bfnm{Xiang}\binits{X.}},
  \bauthor{\bsnm{Sun},~\bfnm{Will~Wei}\binits{W.~W.}},
  \bauthor{\bsnm{Wang},~\bfnm{Zhaoran}\binits{Z.}},
  \bauthor{\bsnm{Liu},~\bfnm{Han}\binits{H.}},
  \bauthor{\bsnm{Yang},~\bfnm{Jian}\binits{J.}} \AND
  \bauthor{\bsnm{Cheng},~\bfnm{Guang}\binits{G.}}
(\byear{2019}).
\btitle{Tensor graphical model: Non-convex optimization and statistical
  inference}.
\bjournal{IEEE transactions on pattern analysis and machine intelligence}
\bvolume{42}
\bpages{2024--2037}.
\end{barticle}
\endbibitem

\bibitem[\protect\citeauthoryear{Manceur and
  Dutilleul}{2013}]{manceur2013maximum}
\begin{barticle}[author]
\bauthor{\bsnm{Manceur},~\bfnm{Ameur~M}\binits{A.~M.}} \AND
  \bauthor{\bsnm{Dutilleul},~\bfnm{Pierre}\binits{P.}}
(\byear{2013}).
\btitle{Maximum likelihood estimation for the tensor normal distribution:
  Algorithm, minimum sample size, and empirical bias and dispersion}.
\bjournal{Journal of Computational and Applied Mathematics}
\bvolume{239}
\bpages{37--49}.
\end{barticle}
\endbibitem

\bibitem[\protect\citeauthoryear{Mardia and Goodall}{1993}]{mardia1993spatial}
\begin{barticle}[author]
\bauthor{\bsnm{Mardia},~\bfnm{Kanti~V}\binits{K.~V.}} \AND
  \bauthor{\bsnm{Goodall},~\bfnm{Colin~R}\binits{C.~R.}}
(\byear{1993}).
\btitle{Spatial-temporal analysis of multivariate environmental monitoring
  data}.
\bjournal{Multivariate environmental statistics}
\bvolume{6}
\bpages{347--385}.
\end{barticle}
\endbibitem

\bibitem[\protect\citeauthoryear{Matthews}{1975}]{matthews1975comparison}
\begin{barticle}[author]
\bauthor{\bsnm{Matthews},~\bfnm{Brian~W}\binits{B.~W.}}
(\byear{1975}).
\btitle{Comparison of the predicted and observed secondary structure of T4
  phage lysozyme}.
\bjournal{Biochimica et Biophysica Acta (BBA)-Protein Structure}
\bvolume{405}
\bpages{442--451}.
\end{barticle}
\endbibitem

\bibitem[\protect\citeauthoryear{Meinshausen and
  B{\"u}hlmann}{2006}]{meinshausen2006high}
\begin{barticle}[author]
\bauthor{\bsnm{Meinshausen},~\bfnm{Nicolai}\binits{N.}} \AND
  \bauthor{\bsnm{B{\"u}hlmann},~\bfnm{Peter}\binits{P.}}
(\byear{2006}).
\btitle{High-dimensional graphs and variable selection with the lasso}.
\bjournal{The annals of statistics}
\bvolume{34}
\bpages{1436--1462}.
\end{barticle}
\endbibitem

\bibitem[\protect\citeauthoryear{Min, Mai and Zhang}{2022}]{min2022fast}
\begin{barticle}[author]
\bauthor{\bsnm{Min},~\bfnm{Keqian}\binits{K.}},
  \bauthor{\bsnm{Mai},~\bfnm{Qing}\binits{Q.}} \AND
  \bauthor{\bsnm{Zhang},~\bfnm{Xin}\binits{X.}}
(\byear{2022}).
\btitle{Fast and separable estimation in high-dimensional tensor Gaussian
  graphical models}.
\bjournal{Journal of Computational and Graphical Statistics}
\bvolume{31}
\bpages{294--300}.
\end{barticle}
\endbibitem

\bibitem[\protect\citeauthoryear{Mo et~al.}{2022}]{mo2022point3d}
\begin{barticle}[author]
\bauthor{\bsnm{Mo},~\bfnm{Shentong}\binits{S.}},
  \bauthor{\bsnm{Xia},~\bfnm{Jingfei}\binits{J.}},
  \bauthor{\bsnm{Tan},~\bfnm{Xiaoqing}\binits{X.}} \AND
  \bauthor{\bsnm{Raj},~\bfnm{Bhiksha}\binits{B.}}
(\byear{2022}).
\btitle{Point3D: tracking actions as moving points with 3D CNNs}.
\bjournal{arXiv preprint arXiv:2203.10584}.
\end{barticle}
\endbibitem

\bibitem[\protect\citeauthoryear{Molstad and
  Rothman}{2019}]{molstad2019penalized}
\begin{barticle}[author]
\bauthor{\bsnm{Molstad},~\bfnm{Aaron~J}\binits{A.~J.}} \AND
  \bauthor{\bsnm{Rothman},~\bfnm{Adam~J}\binits{A.~J.}}
(\byear{2019}).
\btitle{A penalized likelihood method for classification with matrix-valued
  predictors}.
\bjournal{Journal of Computational and Graphical Statistics}
\bvolume{28}
\bpages{11--22}.
\end{barticle}
\endbibitem

\bibitem[\protect\citeauthoryear{Ohlson, Ahmad and
  Von~Rosen}{2013}]{ohlson2013multilinear}
\begin{barticle}[author]
\bauthor{\bsnm{Ohlson},~\bfnm{Martin}\binits{M.}},
  \bauthor{\bsnm{Ahmad},~\bfnm{M~Rauf}\binits{M.~R.}} \AND
  \bauthor{\bsnm{Von~Rosen},~\bfnm{Dietrich}\binits{D.}}
(\byear{2013}).
\btitle{The multilinear normal distribution: Introduction and some basic
  properties}.
\bjournal{Journal of Multivariate Analysis}
\bvolume{113}
\bpages{37--47}.
\end{barticle}
\endbibitem

\bibitem[\protect\citeauthoryear{Perkel}{2019}]{perkel2019julia}
\begin{barticle}[author]
\bauthor{\bsnm{Perkel},~\bfnm{Jeffrey~M}\binits{J.~M.}}
(\byear{2019}).
\btitle{Julia: come for the syntax, stay for the speed}.
\bjournal{Nature}
\bvolume{572}
\bpages{141--143}.
\end{barticle}
\endbibitem

\bibitem[\protect\citeauthoryear{Pourahmadi}{2011}]{pourahmadi2011covariance}
\begin{barticle}[author]
\bauthor{\bsnm{Pourahmadi},~\bfnm{Mohsen}\binits{M.}}
(\byear{2011}).
\btitle{Covariance estimation: The GLM and regularization perspectives}.
\bjournal{Statistical Science}
\bvolume{26}
\bpages{369--387}.
\end{barticle}
\endbibitem

\bibitem[\protect\citeauthoryear{Pouryazdian, Beheshti and
  Krishnan}{2016}]{pouryazdian2016candecomp}
\begin{barticle}[author]
\bauthor{\bsnm{Pouryazdian},~\bfnm{Saeed}\binits{S.}},
  \bauthor{\bsnm{Beheshti},~\bfnm{Soosan}\binits{S.}} \AND
  \bauthor{\bsnm{Krishnan},~\bfnm{Sridhar}\binits{S.}}
(\byear{2016}).
\btitle{CANDECOMP/PARAFAC model order selection based on reconstruction error
  in the presence of kronecker structured colored noise}.
\bjournal{Digital Signal Processing}
\bvolume{48}
\bpages{12--26}.
\end{barticle}
\endbibitem

\bibitem[\protect\citeauthoryear{Rabusseau and Kadri}{2016}]{rabusseau2016low}
\begin{barticle}[author]
\bauthor{\bsnm{Rabusseau},~\bfnm{Guillaume}\binits{G.}} \AND
  \bauthor{\bsnm{Kadri},~\bfnm{Hachem}\binits{H.}}
(\byear{2016}).
\btitle{Low-rank regression with tensor responses}.
\bjournal{Advances in Neural Information Processing Systems}
\bvolume{29}
\bpages{1867--1875}.
\end{barticle}
\endbibitem

\bibitem[\protect\citeauthoryear{Richard and
  Montanari}{2014}]{richard2014statistical}
\begin{barticle}[author]
\bauthor{\bsnm{Richard},~\bfnm{Emile}\binits{E.}} \AND
  \bauthor{\bsnm{Montanari},~\bfnm{Andrea}\binits{A.}}
(\byear{2014}).
\btitle{A statistical model for tensor PCA}.
\bjournal{Advances in Neural Information Processing Systems}
\bvolume{27}
\bpages{2897--2905}.
\end{barticle}
\endbibitem

\bibitem[\protect\citeauthoryear{Roberts}{2006}]{roberts2006slow}
\begin{barticle}[author]
\bauthor{\bsnm{Roberts},~\bfnm{B}\binits{B.}}
(\byear{2006}).
\btitle{Slow magnetohydrodynamic waves in the solar atmosphere}.
\bjournal{Philosophical Transactions of the Royal Society A: Mathematical,
  Physical and Engineering Sciences}
\bvolume{364}
\bpages{447--460}.
\end{barticle}
\endbibitem

\bibitem[\protect\citeauthoryear{Stocker}{2011}]{stocker2011introduction}
\begin{bbook}[author]
\bauthor{\bsnm{Stocker},~\bfnm{Thomas}\binits{T.}}
(\byear{2011}).
\btitle{Introduction to climate modelling}.
\bpublisher{Springer Science \& Business Media}.
\end{bbook}
\endbibitem

\bibitem[\protect\citeauthoryear{Strobach}{1995}]{strobach1995low}
\begin{barticle}[author]
\bauthor{\bsnm{Strobach},~\bfnm{Peter}\binits{P.}}
(\byear{1995}).
\btitle{Low-rank detection of multichannel Gaussian signals using block matrix
  approximation}.
\bjournal{IEEE Transactions on Signal Processing}
\bvolume{43}
\bpages{233--242}.
\end{barticle}
\endbibitem

\bibitem[\protect\citeauthoryear{Sun, Hao and Li}{}]{suntensors}
\begin{barticle}[author]
\bauthor{\bsnm{Sun},~\bfnm{Will~Wei}\binits{W.~W.}},
  \bauthor{\bsnm{Hao},~\bfnm{Botao}\binits{B.}} \AND
  \bauthor{\bsnm{Li},~\bfnm{Lexin}\binits{L.}}
\btitle{Tensors in Modern Statistical Learning}.
\end{barticle}
\endbibitem

\bibitem[\protect\citeauthoryear{Sun et~al.}{2021}]{sun2021predicting}
\begin{barticle}[author]
\bauthor{\bsnm{Sun},~\bfnm{Zeyu}\binits{Z.}},
  \bauthor{\bsnm{Bobra},~\bfnm{Monica}\binits{M.}},
  \bauthor{\bsnm{Wang},~\bfnm{Xiantong}\binits{X.}},
  \bauthor{\bsnm{Wang},~\bfnm{Yu}\binits{Y.}},
  \bauthor{\bsnm{Sun},~\bfnm{Hu}\binits{H.}},
  \bauthor{\bsnm{Gombosi},~\bfnm{Tamas}\binits{T.}},
  \bauthor{\bsnm{Chen},~\bfnm{Yang}\binits{Y.}} \AND
  \bauthor{\bsnm{Hero},~\bfnm{Alfred}\binits{A.}}
(\byear{2021}).
\btitle{Predicting Solar Flares using CNN and LSTM on Two Solar Cycles of
  Active Region Data}.
\bjournal{Earth and Space Science Open Archive}
\bpages{32}.
\bdoi{10.1002/essoar.10508256.1}
\end{barticle}
\endbibitem

\bibitem[\protect\citeauthoryear{Tan et~al.}{2018}]{tan2018changepoint}
\begin{binproceedings}[author]
\bauthor{\bsnm{Tan},~\bfnm{Xiaoqing}\binits{X.}},
  \bauthor{\bsnm{Ross},~\bfnm{Christopher~A}\binits{C.~A.}},
  \bauthor{\bsnm{Miller},~\bfnm{Michaeli}\binits{M.}} \AND
  \bauthor{\bsnm{Tang},~\bfnm{Xiaoying}\binits{X.}}
(\byear{2018}).
\btitle{Changepoint analysis of putamen and thalamus subregions in premanifest
  huntington's disease}.
In \bbooktitle{2018 IEEE 15th International Symposium on Biomedical Imaging
  (ISBI 2018)}
\bpages{531--535}.
\bpublisher{IEEE}.
\end{binproceedings}
\endbibitem

\bibitem[\protect\citeauthoryear{Tan et~al.}{2022a}]{tan2022rise}
\begin{barticle}[author]
\bauthor{\bsnm{Tan},~\bfnm{Xiaoqing}\binits{X.}},
  \bauthor{\bsnm{Qi},~\bfnm{Zhengling}\binits{Z.}},
  \bauthor{\bsnm{Seymour},~\bfnm{Christopher~W}\binits{C.~W.}} \AND
  \bauthor{\bsnm{Tang},~\bfnm{Lu}\binits{L.}}
(\byear{2022}a).
\btitle{RISE: Robust Individualized Decision Learning with Sensitive
  Variables}.
\bjournal{arXiv preprint arXiv:2211.06569}.
\end{barticle}
\endbibitem

\bibitem[\protect\citeauthoryear{Tan et~al.}{2022b}]{tan2022identifying}
\begin{binproceedings}[author]
\bauthor{\bsnm{Tan},~\bfnm{Xiaoqing}\binits{X.}},
  \bauthor{\bsnm{Abberbock},~\bfnm{Judah}\binits{J.}},
  \bauthor{\bsnm{Rastogi},~\bfnm{Priya}\binits{P.}} \AND
  \bauthor{\bsnm{Tang},~\bfnm{Gong}\binits{G.}}
(\byear{2022}b).
\btitle{Identifying Principal Stratum Causal Effects Conditional on a
  Post-treatment Intermediate Response}.
In \bbooktitle{Conference on Causal Learning and Reasoning}
\bpages{734--753}.
\bpublisher{PMLR}.
\end{binproceedings}
\endbibitem

\bibitem[\protect\citeauthoryear{Tan et~al.}{2022c}]{tan2022doubly}
\begin{barticle}[author]
\bauthor{\bsnm{Tan},~\bfnm{Xiaoqing}\binits{X.}},
  \bauthor{\bsnm{Yang},~\bfnm{Shu}\binits{S.}},
  \bauthor{\bsnm{Ye},~\bfnm{Wenyu}\binits{W.}},
  \bauthor{\bsnm{Faries},~\bfnm{Douglas~E}\binits{D.~E.}},
  \bauthor{\bsnm{Lipkovich},~\bfnm{Ilya}\binits{I.}} \AND
  \bauthor{\bsnm{Kadziola},~\bfnm{Zbigniew}\binits{Z.}}
(\byear{2022}c).
\btitle{When Doubly Robust Methods Meet Machine Learning for Estimating
  Treatment Effects from Real-World Data: A Comparative Study}.
\bjournal{arXiv preprint arXiv:2204.10969}.
\end{barticle}
\endbibitem

\bibitem[\protect\citeauthoryear{Tan et~al.}{2022d}]{tan2022tree}
\begin{binproceedings}[author]
\bauthor{\bsnm{Tan},~\bfnm{Xiaoqing}\binits{X.}},
  \bauthor{\bsnm{Chang},~\bfnm{Chung-Chou~H}\binits{C.-C.~H.}},
  \bauthor{\bsnm{Zhou},~\bfnm{Ling}\binits{L.}} \AND
  \bauthor{\bsnm{Tang},~\bfnm{Lu}\binits{L.}}
(\byear{2022}d).
\btitle{A tree-based model averaging approach for personalized treatment effect
  estimation from heterogeneous data sources}.
In \bbooktitle{International Conference on Machine Learning}
\bpages{21013--21036}.
\bpublisher{PMLR}.
\end{binproceedings}
\endbibitem

\bibitem[\protect\citeauthoryear{Teng and Huang}{2009}]{teng2009statistical}
\begin{barticle}[author]
\bauthor{\bsnm{Teng},~\bfnm{Siew~Leng}\binits{S.~L.}} \AND
  \bauthor{\bsnm{Huang},~\bfnm{Haiyan}\binits{H.}}
(\byear{2009}).
\btitle{A statistical framework to infer functional gene relationships from
  biologically interrelated microarray experiments}.
\bjournal{Journal of the American Statistical Association}
\bvolume{104}
\bpages{465--473}.
\end{barticle}
\endbibitem

\bibitem[\protect\citeauthoryear{Tsiligkaridis and
  Hero}{2013}]{tsiligkaridis2013covariance}
\begin{barticle}[author]
\bauthor{\bsnm{Tsiligkaridis},~\bfnm{Theodoros}\binits{T.}} \AND
  \bauthor{\bsnm{Hero},~\bfnm{Alfred~O}\binits{A.~O.}}
(\byear{2013}).
\btitle{Covariance estimation in high dimensions via kronecker product
  expansions}.
\bjournal{IEEE Transactions on Signal Processing}
\bvolume{61}
\bpages{5347--5360}.
\end{barticle}
\endbibitem

\bibitem[\protect\citeauthoryear{Tsiligkaridis, Hero and
  Zhou}{2013}]{tsiligkaridis2013convergence}
\begin{barticle}[author]
\bauthor{\bsnm{Tsiligkaridis},~\bfnm{Theodoros}\binits{T.}},
  \bauthor{\bsnm{Hero},~\bfnm{Alfred~O}\binits{A.~O.}} \AND
  \bauthor{\bsnm{Zhou},~\bfnm{Shuheng}\binits{S.}}
(\byear{2013}).
\btitle{On convergence of kronecker graphical lasso algorithms}.
\bjournal{IEEE transactions on signal processing}
\bvolume{61}
\bpages{1743--1755}.
\end{barticle}
\endbibitem

\bibitem[\protect\citeauthoryear{van Driel-Gesztelyi and
  Green}{2015}]{van2015evolution}
\begin{barticle}[author]
\bauthor{\bparticle{van} \bsnm{Driel-Gesztelyi},~\bfnm{Lidia}\binits{L.}} \AND
  \bauthor{\bsnm{Green},~\bfnm{Lucie~May}\binits{L.~M.}}
(\byear{2015}).
\btitle{Evolution of active regions}.
\bjournal{Living Reviews in Solar Physics}
\bvolume{12}
\bpages{1--98}.
\end{barticle}
\endbibitem

\bibitem[\protect\citeauthoryear{Van~Loan and
  Pitsianis}{1993}]{van1993approximation}
\begin{bincollection}[author]
\bauthor{\bsnm{Van~Loan},~\bfnm{Charles~F}\binits{C.~F.}} \AND
  \bauthor{\bsnm{Pitsianis},~\bfnm{Nikos}\binits{N.}}
(\byear{1993}).
\btitle{Approximation with Kronecker products}.
In \bbooktitle{Linear algebra for large scale and real-time applications}
\bpages{293--314}.
\bpublisher{Springer}.
\end{bincollection}
\endbibitem

\bibitem[\protect\citeauthoryear{Varin, Reid and
  Firth}{2011}]{varin2011overview}
\begin{barticle}[author]
\bauthor{\bsnm{Varin},~\bfnm{Cristiano}\binits{C.}},
  \bauthor{\bsnm{Reid},~\bfnm{Nancy}\binits{N.}} \AND
  \bauthor{\bsnm{Firth},~\bfnm{David}\binits{D.}}
(\byear{2011}).
\btitle{An overview of composite likelihood methods}.
\bjournal{Statistica Sinica}
\bpages{5--42}.
\end{barticle}
\endbibitem

\bibitem[\protect\citeauthoryear{Wang and Hero}{2021a}]{wang2021multiway}
\begin{barticle}[author]
\bauthor{\bsnm{Wang},~\bfnm{Yu}\binits{Y.}} \AND
  \bauthor{\bsnm{Hero},~\bfnm{Alfred}\binits{A.}}
(\byear{2021}a).
\btitle{Multiway Ensemble Kalman Filter}.
\bjournal{arXiv preprint arXiv:2112.04322}.
\end{barticle}
\endbibitem

\bibitem[\protect\citeauthoryear{Wang and Hero}{2021b}]{wang2021sg}
\begin{barticle}[author]
\bauthor{\bsnm{Wang},~\bfnm{Yu}\binits{Y.}} \AND
  \bauthor{\bsnm{Hero},~\bfnm{Alfred}\binits{A.}}
(\byear{2021}b).
\btitle{SG-PALM: a Fast Physically Interpretable Tensor Graphical Model}.
\bjournal{International Conference on Machine Learning (ICML), arXiv preprint
  arXiv:2105.12271}.
\end{barticle}
\endbibitem

\bibitem[\protect\citeauthoryear{Wang, Jang and Hero}{2020}]{wang2020sylvester}
\begin{binproceedings}[author]
\bauthor{\bsnm{Wang},~\bfnm{Yu}\binits{Y.}},
  \bauthor{\bsnm{Jang},~\bfnm{Byoungwook}\binits{B.}} \AND
  \bauthor{\bsnm{Hero},~\bfnm{Alfred}\binits{A.}}
(\byear{2020}).
\btitle{The sylvester graphical lasso (syglasso)}.
In \bbooktitle{International Conference on Artificial Intelligence and
  Statistics}
\bpages{1943--1953}.
\bpublisher{PMLR}.
\end{binproceedings}
\endbibitem

\bibitem[\protect\citeauthoryear{Wang, Sun and Hero}{2022}]{WANG2022100308}
\begin{barticle}[author]
\bauthor{\bsnm{Wang},~\bfnm{Yu}\binits{Y.}},
  \bauthor{\bsnm{Sun},~\bfnm{Zeyu}\binits{Z.}} \AND
  \bauthor{\bsnm{Hero},~\bfnm{Alfred}\binits{A.}}
(\byear{2022}).
\btitle{TensorGraphicalModels: A Julia toolbox for multiway covariance models
  and ensemble Kalman filter}.
\bjournal{Software Impacts}
\bvolume{13}
\bpages{100308}.
\bdoi{https://doi.org/10.1016/j.simpa.2022.100308}
\end{barticle}
\endbibitem

\bibitem[\protect\citeauthoryear{Wang et~al.}{2020}]{wang2020predicting}
\begin{barticle}[author]
\bauthor{\bsnm{Wang},~\bfnm{Xiantong}\binits{X.}},
  \bauthor{\bsnm{Chen},~\bfnm{Yang}\binits{Y.}},
  \bauthor{\bsnm{Toth},~\bfnm{Gabor}\binits{G.}},
  \bauthor{\bsnm{Manchester},~\bfnm{Ward~B}\binits{W.~B.}},
  \bauthor{\bsnm{Gombosi},~\bfnm{Tamas~I}\binits{T.~I.}},
  \bauthor{\bsnm{Hero},~\bfnm{Alfred~O}\binits{A.~O.}},
  \bauthor{\bsnm{Jiao},~\bfnm{Zhenbang}\binits{Z.}},
  \bauthor{\bsnm{Sun},~\bfnm{Hu}\binits{H.}},
  \bauthor{\bsnm{Jin},~\bfnm{Meng}\binits{M.}} \AND
  \bauthor{\bsnm{Liu},~\bfnm{Yang}\binits{Y.}}
(\byear{2020}).
\btitle{Predicting solar flares with machine learning: investigating solar
  cycle dependence}.
\bjournal{The Astrophysical Journal}
\bvolume{895}
\bpages{3}.
\end{barticle}
\endbibitem

\bibitem[\protect\citeauthoryear{Wei and Minsker}{2017}]{wei2017heavytailed}
\begin{binproceedings}[author]
\bauthor{\bsnm{Wei},~\bfnm{Xiaohan}\binits{X.}} \AND
  \bauthor{\bsnm{Minsker},~\bfnm{Stanislav}\binits{S.}}
(\byear{2017}).
\btitle{Estimation of the covariance structure of heavy-tailed distributions}.
In \bbooktitle{Advances in Neural Information Processing Systems}
(\beditor{\bfnm{I.}\binits{I.}~\bsnm{Guyon}},
  \beditor{\bfnm{U.~V.}\binits{U.~V.}~\bsnm{Luxburg}},
  \beditor{\bfnm{S.}\binits{S.}~\bsnm{Bengio}},
  \beditor{\bfnm{H.}\binits{H.}~\bsnm{Wallach}},
  \beditor{\bfnm{R.}\binits{R.}~\bsnm{Fergus}},
  \beditor{\bfnm{S.}\binits{S.}~\bsnm{Vishwanathan}} \AND
  \beditor{\bfnm{R.}\binits{R.}~\bsnm{Garnett}}, eds.)
\bvolume{30}.
\bpublisher{Curran Associates, Inc.}
\end{binproceedings}
\endbibitem

\bibitem[\protect\citeauthoryear{Werner, Jansson and
  Stoica}{2008}]{werner2008estimation}
\begin{barticle}[author]
\bauthor{\bsnm{Werner},~\bfnm{Karl}\binits{K.}},
  \bauthor{\bsnm{Jansson},~\bfnm{Magnus}\binits{M.}} \AND
  \bauthor{\bsnm{Stoica},~\bfnm{Petre}\binits{P.}}
(\byear{2008}).
\btitle{On estimation of covariance matrices with Kronecker product structure}.
\bjournal{IEEE Transactions on Signal Processing}
\bvolume{56}
\bpages{478--491}.
\end{barticle}
\endbibitem

\bibitem[\protect\citeauthoryear{Wiesel et~al.}{2015}]{wiesel2015structured}
\begin{barticle}[author]
\bauthor{\bsnm{Wiesel},~\bfnm{Ami}\binits{A.}},
  \bauthor{\bsnm{Zhang},~\bfnm{Teng}\binits{T.}} \betal{et~al.}
(\byear{2015}).
\btitle{Structured robust covariance estimation}.
\bjournal{Foundations and Trends{\textregistered} in Signal Processing}
\bvolume{8}
\bpages{127--216}.
\end{barticle}
\endbibitem

\bibitem[\protect\citeauthoryear{Xu, Zhang and Gu}{2017}]{xu2017efficient}
\begin{binproceedings}[author]
\bauthor{\bsnm{Xu},~\bfnm{Pan}\binits{P.}},
  \bauthor{\bsnm{Zhang},~\bfnm{Tingting}\binits{T.}} \AND
  \bauthor{\bsnm{Gu},~\bfnm{Quanquan}\binits{Q.}}
(\byear{2017}).
\btitle{Efficient algorithm for sparse tensor-variate gaussian graphical models
  via gradient descent}.
In \bbooktitle{Artificial Intelligence and Statistics}
\bpages{923--932}.
\bpublisher{PMLR}.
\end{binproceedings}
\endbibitem

\bibitem[\protect\citeauthoryear{Yin and Li}{2012}]{yin2012model}
\begin{barticle}[author]
\bauthor{\bsnm{Yin},~\bfnm{Jianxin}\binits{J.}} \AND
  \bauthor{\bsnm{Li},~\bfnm{Hongzhe}\binits{H.}}
(\byear{2012}).
\btitle{Model selection and estimation in the matrix normal graphical model}.
\bjournal{Journal of multivariate analysis}
\bvolume{107}
\bpages{119--140}.
\end{barticle}
\endbibitem

\bibitem[\protect\citeauthoryear{Yoon and Kim}{2020}]{yoon2020eiglasso}
\begin{binproceedings}[author]
\bauthor{\bsnm{Yoon},~\bfnm{Jun~Ho}\binits{J.~H.}} \AND
  \bauthor{\bsnm{Kim},~\bfnm{Seyoung}\binits{S.}}
(\byear{2020}).
\btitle{EiGLasso: Scalable estimation of Cartesian product of sparse inverse
  covariance matrices}.
In \bbooktitle{Conference on Uncertainty in Artificial Intelligence}
\bpages{1248--1257}.
\bpublisher{PMLR}.
\end{binproceedings}
\endbibitem

\bibitem[\protect\citeauthoryear{Yu et~al.}{2001}]{yu2001second}
\begin{binproceedings}[author]
\bauthor{\bsnm{Yu},~\bfnm{Kai}\binits{K.}},
  \bauthor{\bsnm{Bengtsson},~\bfnm{Mats}\binits{M.}},
  \bauthor{\bsnm{Ottersten},~\bfnm{Bj{\"o}rn}\binits{B.}},
  \bauthor{\bsnm{McNamara},~\bfnm{Darren}\binits{D.}},
  \bauthor{\bsnm{Karlsson},~\bfnm{Peter}\binits{P.}} \AND
  \bauthor{\bsnm{Beach},~\bfnm{Mark}\binits{M.}}
(\byear{2001}).
\btitle{Second order statistics of NLOS indoor MIMO channels based on 5.2 GHz
  measurements}.
In \bbooktitle{GLOBECOM'01. IEEE Global Telecommunications Conference (Cat. No.
  01CH37270)}
\bvolume{1}
\bpages{156--160}.
\bpublisher{IEEE}.
\end{binproceedings}
\endbibitem

\bibitem[\protect\citeauthoryear{Yuan and Lin}{2007}]{yuan2007model}
\begin{barticle}[author]
\bauthor{\bsnm{Yuan},~\bfnm{Ming}\binits{M.}} \AND
  \bauthor{\bsnm{Lin},~\bfnm{Yi}\binits{Y.}}
(\byear{2007}).
\btitle{Model selection and estimation in the Gaussian graphical model}.
\bjournal{Biometrika}
\bvolume{94}
\bpages{19--35}.
\end{barticle}
\endbibitem

\bibitem[\protect\citeauthoryear{Zhang and Schneider}{2010}]{zhang2010learning}
\begin{binproceedings}[author]
\bauthor{\bsnm{Zhang},~\bfnm{Yi}\binits{Y.}} \AND
  \bauthor{\bsnm{Schneider},~\bfnm{Jeff~G}\binits{J.~G.}}
(\byear{2010}).
\btitle{Learning multiple tasks with a sparse matrix-normal penalty.}
In \bbooktitle{NIPS}
\bvolume{6}
\bpages{2}.
\end{binproceedings}
\endbibitem

\bibitem[\protect\citeauthoryear{Zhang and Xia}{2018}]{zhang2018tensor}
\begin{barticle}[author]
\bauthor{\bsnm{Zhang},~\bfnm{Anru}\binits{A.}} \AND
  \bauthor{\bsnm{Xia},~\bfnm{Dong}\binits{D.}}
(\byear{2018}).
\btitle{Tensor SVD: Statistical and computational limits}.
\bjournal{IEEE Transactions on Information Theory}
\bvolume{64}
\bpages{7311--7338}.
\end{barticle}
\endbibitem

\bibitem[\protect\citeauthoryear{Zhou}{2014}]{zhou2014gemini}
\begin{barticle}[author]
\bauthor{\bsnm{Zhou},~\bfnm{Shuheng}\binits{S.}}
(\byear{2014}).
\btitle{Gemini: Graph estimation with matrix variate normal instances}.
\bjournal{The Annals of Statistics}
\bvolume{42}
\bpages{532--562}.
\end{barticle}
\endbibitem

\bibitem[\protect\citeauthoryear{Zhou, Li and Zhu}{2013}]{zhou2013tensor}
\begin{barticle}[author]
\bauthor{\bsnm{Zhou},~\bfnm{Hua}\binits{H.}},
  \bauthor{\bsnm{Li},~\bfnm{Lexin}\binits{L.}} \AND
  \bauthor{\bsnm{Zhu},~\bfnm{Hongtu}\binits{H.}}
(\byear{2013}).
\btitle{Tensor regression with applications in neuroimaging data analysis}.
\bjournal{Journal of the American Statistical Association}
\bvolume{108}
\bpages{540--552}.
\end{barticle}
\endbibitem

\bibitem[\protect\citeauthoryear{Zhou et~al.}{2021}]{zhou2021partially}
\begin{barticle}[author]
\bauthor{\bsnm{Zhou},~\bfnm{Jie}\binits{J.}},
  \bauthor{\bsnm{Sun},~\bfnm{Will~Wei}\binits{W.~W.}},
  \bauthor{\bsnm{Zhang},~\bfnm{Jingfei}\binits{J.}} \AND
  \bauthor{\bsnm{Li},~\bfnm{Lexin}\binits{L.}}
(\byear{2021}).
\btitle{Partially observed dynamic tensor response regression}.
\bjournal{Journal of the American Statistical Association}
\bpages{1--16}.
\end{barticle}
\endbibitem

\end{thebibliography}


\end{document}